\documentstyle[prl,aps,epsf,twocolumn]{revtex}

\newcommand{\la}{\langle}
\newcommand{\ra}{\rangle}
\newcommand{\beq}{\begin{eqnarray}}
\newcommand{\eeq}{\end{eqnarray}}

\newcommand{\btem}{\bibitem}
\newcommand{\vpi}{\vec{\pi}}
\newcommand{\bep}{{1 \over \bar{\varepsilon}}}
\newcommand{\vep}{\varepsilon}

\newcommand{\ms}{m_{0\sigma}}
\newcommand{\mh}{m_{0\phi}}
\newcommand{\mev}{\mbox{MeV}}
\renewcommand{\mp}{m_{0\pi}}
\renewcommand{\d}{\partial}

\begin{document}

%\preprint{UTHEP-379, Feb. 1998}
\draft

\title{Optimized Perturbation Theory at Finite Temperature} 

\author{S. Chiku$^{(1)}$ and T. Hatsuda$^{(2)}$}
\address{Institute of Physics, University of Tsukuba,
 Tsukuba, Ibaraki 305, Japan}
\address{$^{(1)}$ Yukawa Institute for Theoretical Physics,
Kyoto University, Kyoto 606, Japan}
\address{$^{(2)}$ Physics Department,
Kyoto University, Kyoto 606, Japan}

\date{\today}
 
\maketitle

\begin{abstract}
 An optimized perturbation theory (OPT) at finite temperature $T$, 
 which resums 
 higher order terms in the naive perturbation, is developed
 in $O(N)$ $\phi^4$ theory.
 It is proved  that (i) the renormalization of the ultra-violet
 divergences can be carried out systematically in any given order of OPT,
 and (ii)  the Nambu-Goldstone theorem is fulfilled for arbitrary $N$ 
 and for any given order of OPT.
 The method is applied for the $O(4)$ $\sigma$ model 
 to study the
 soft modes associated with the chiral transition in quantum chromodynamics.
 Threshold enhancement of the spectral functions at finite $T$
 in the scalar and pseudo-scalar channels
 is shown to be a typical signal of the chiral transition.

\end{abstract}


%\narrowtext
\section{Introduction}

 One of the main goals of the ultra-relativistic
 heavy-ion experiments planned
 at RHIC and LHC \cite{QM96} is to observe the structural change of 
 the ground state of quantum chromodynamics (QCD)
 at finite temperature ($T$), namely
 the phase transition to the quark-gluon plasma.
 The numerical simulation based on the lattice QCD
 is  a powerful tool to study the static nature of this 
 phase transition, in which the critical temperature and 
 the critical exponents 
 are actively studied \cite{LAT96}.
 In particular, there exit numerical evidences that the chiral transition
 for massless two flavors is of 
 second order, although the case for 
 the real world (two light quarks + one medium-heavy quark)
 is not settled yet \cite{ukawa}.

 If the phase transition is of second order or is close to it, 
 there arises long range fluctuations
 in both spatial and temporal (real-time) directions. 
 The latter is usually called the soft mode  and has been used as
 a probe to study phase transitions 
 of solid states and condensed matters
 \cite{GK}.

 Despite the experimental significance of the soft modes in QCD,
 the lattice QCD simulations cannot treat such real-time modes
 in a straightforward manner.  This is why effective theories of QCD
 have been adopted to study the time-dependent  phenomena 
(see the reviews  \cite{HK94,RAJ}
 and references cited therein.) However, 
 even in tractable effective theories such as the 
 linear $\sigma$-model, 
 there exist subtleties at finite $T$.
 In fact, the necessity of the resummation of 
 higher order terms in perturbative expansions both at high $T$
 and low $T$ has been known for a long time \cite{WEIN74,KL76}.
 Also, the renormalization of the ultra-violet (UV)
 divergences and related issue in resumed perturbation theories
 have been discussed in the literatures especially for theories with
 spontaneous symmetry breaking (SSB) \cite{renorm}.

 Recently, we have reported our
 analysis of a particular resummation method and 
 its application to the soft modes in QCD \cite{CH98}.
 The present paper contains not only the detailed description of our previous
 analysis but also further investigations.

 The purpose of the present paper is twofold.
 Firstly, we will develop an improved loop-wise expansion
 at finite $T$. Our starting point is 
 the optimized perturbation theory (OPT)
 (or sometimes called  
 delta-expansion, variational perturbation
 theory etc)
 which is a generalization of the mean-field method \cite{NJL}
 and is known to work in various quantum systems \cite{OPT}.
 Its application to field theory at finite $T$ has been 
 considered in ref.\cite{oko,BM} for the first time.
 We will further develop the idea  
 and  prove the renormalizability and 
 the Nambu-Goldstone (NG) theorem in
 $O(N)$ $\phi^4$ theory  at finite $T$ 
 order by order in OPT.
 Our second purpose is
 to study the soft modes  associated with the
 chiral transition in QCD 
 by taking into account interactions among the soft modes (mode 
 couplings).
 The use of OPT is essential for this purpose, which 
 will be demonstrated using  the $O(4)$ $\sigma$ model.
 
 The organization of this paper is as follows.
 In section II,  
 we introduce a loop-wise expansion 
 on the basis of  OPT. 
 The renormalization of UV divergences and  the realization of the
  NG theorem in this method are also discussed.
 In section III, 
 we will apply OPT developed in section II
 for the $O(4)$ $\sigma$ model to study the 
 spectral functions of the $\sigma$-meson  and the $\pi$-meson
 at $T \neq 0$. It is also examined the 
 detectability of the soft modes
 by the diphoton process 
 $\sigma \rightarrow 2 \gamma$ in hot hadronic matter.
 Section IV is devoted to summary and concluding remarks.

\section{Optimized Perturbation at $T \neq 0$}
\label{RPT}

\subsection{Necessity of resummation at finite $T$}

 It has been known that naive  perturbations either by a
 coupling constant or by number of loops  break down
 at $T \neq 0$, and proper resummation of higher orders
 is necessary \cite{WEIN74}.
 In fact, no matter what a small dimensionless coupling (say $\lambda$)
 seems to control the perturbative expansion, the powers of $T$
 compensate the powers of $\lambda$, which invalidates the naive
 expansion.  

\begin{figure}[h]
%  \begin{center}
\centerline{
    \epsfxsize=8.9cm 
    \epsfbox{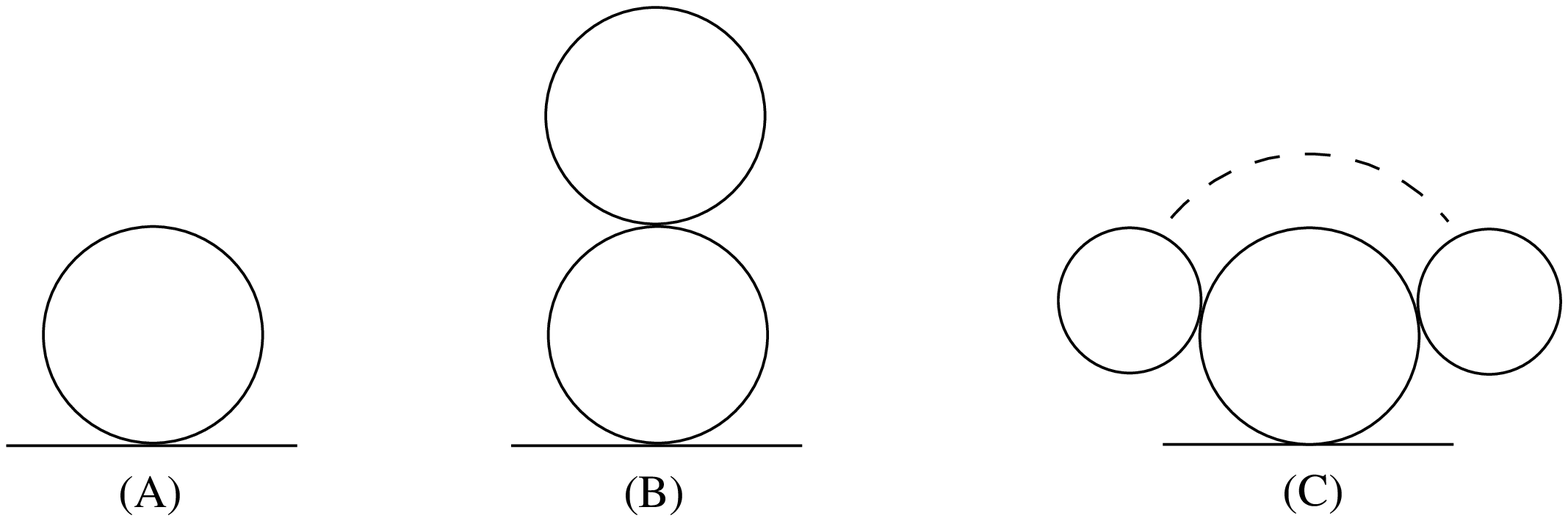}
}
%  \end{center}
     \caption{Bubble and cactus diagrams.}
 \label{tado}
\end{figure}

 This is easily  illustrated  in the $\phi^4$ theory:
\beq
\label{toy}
  {\cal L}  =  {1 \over 2} [(\d \phi)^2 - \mu^2
  \phi^2] -{ \lambda \over 4! }\phi^4 .
\eeq
 Let us  first consider the case $\mu^2 > 0$.
 The lowest order self-energy  diagram
 Fig.\ref{tado} (A) is  $O( \lambda T^2)$ at high $T$. 
 However, Fig.\ref{tado} (B) is
 $O(\lambda T^2 \times {\lambda T \over \mu})$. Furthermore, 
 higher  powers of $T/\mu$ arise in 
 higher loops; e.g. the n-loop diagram
  in Fig.\ref{tado} (C)  is 
 $O(\lambda^n T^{2n-1}/\mu^{2n-3})$.  Thus, one should at least
 resum cactus diagrams
 to get sensible results  at high $T$  \cite{WEIN74,fen}.
 Physics behind this resummation is of course the Debye screening 
 mass in the hot plasma.  

 The naive loop-expansion breaks down also for
 $\mu^2 < 0$.
 The tree-level mass $m_0$ in this case is defined as
\beq  
 m_0^2 =  \mu^2 + {\lambda \over 2} \xi^2(T) ,
\eeq
 where $\xi(T) $ is the thermal expectation value of $\phi$.
 Since $\mu^2$ is negative  and 
 $\xi^2$ decreases as $T$ increases, 
 $m_0^2$ becomes tachyonic even
 below the critical temperature  $T_c$. Therefore, 
 the  naive loop-expansion using the tree-level
 propagator ceases to work even 
 before the symmetry restoration  takes place, and  
  proper resummation of higher loop diagrams is 
 necessary \cite{KL76}.
 Note that, for $T< T_c$, there
 is no reason to believe that only the cactus diagrams 
 shown in Fig.1 are dominant;
 there exists a three-point vertex $\lambda \xi  \phi^3$ which is not
 negligible for $T \sim \xi (T)$.

\subsection{Resummation method}

 For theories without SSB, 
 a systematic resummation method to obtain a sensible
 `` weak-coupling'' expansion at high $T$ was formulated 
 and applied 
 to gauge theory and $\phi^4$ theory successfully \cite{BP}.

 For theories with SSB, however,  loop-expansion rather than the 
 weak-coupling expansion is 
 relevant, since one needs to treat the thermal
 effective potential or the Gibbs free energy.
 We find that the optimized perturbation theory (OPT), 
 which was applied to finite $T$ system in \cite{oko,BM}, 
 can be formulated in such a way that 
 an improved loop-expansion is carried out systematically.
 Also, the method leads to a transparent 
 renormalization procedure and guarantees the
 Nambu-Goldstone theorem order by order
 in the improved loop-expansion.

 In the following, we divide our resummation procedure into three steps
 and apply it to the $\phi^4$ theory.
 The case for  $O(N)$ $\phi^4$ theory will be discussed
 in  Section II.E.

\noindent
{\bf Step 1}: \\
 Start with a  renormalized Lagrangian  with counter terms
\beq
\label{ala1}
  {\cal L}(\phi;\mu^2) 
            & = & 
  {1 \over 2} [(\d \phi)^2 - \mu^2
  \phi^2] -{ \lambda \over 4! }\phi^4   \nonumber \\
& & +{1 \over 2}A(\d \phi)^2
    -{1 \over 2} B \mu^2 \phi^2  -{ \lambda \over 4! }C  \phi^4 
    + D \mu^4 .
\eeq
 Here we have explicitly written the argument $\mu^2$
 in ${\cal L}$ for later use.
 The mass independent renormalization scheme with the 
 dimensional regularization   is assumed in (\ref{ala1}).
  Just for notational simplicity,
  the factor $\kappa^{(4-n)}$ to be
 multiplied to $\lambda$ is omitted ($\kappa$ is the
renormalization
 point and $n$ is the number of dimensions).
  In the actual calculations below, we take the 
 modified minimal subtraction ($\overline{MS}$) scheme.

 The  $c$-number counter term $D\mu^4$,
 which was not considered in \cite{BM},
 is necessary  to make the thermal effective potential
 finite.  Also, it plays a crucial role for
 renormalization in OPT as will be shown in Sec.II.D.

 The thermal effective action
 $\Gamma[\varphi^2]$ is written as the Euclidean
 functional integral \cite{lexp}
\beq
\label{naive-l}
  {\Gamma[\varphi^2]} = \ln  \int[d\phi]
  \exp \left[  {1 \over \delta} \int_0^{1/T} d^4 x
  \left[ {\cal L} (\phi + \varphi;\mu^2) + J \phi
  \right] \right] ,	
\eeq
where $J \equiv - \d \Gamma[\varphi] /\d \varphi $ and
$\int_0^{1/T} d^4 x \equiv \int_0^{1/T} d\tau \int d^3 x $.
 The ``naive'' loop expansion at $T\neq 0$ 
 is defined as an expansion by $\delta$  \cite{lexp2}
 with the  tree-level mass 
 $\mu^2 + \lambda \varphi^2/2$.
 
  Under the naive loop-expansion with (\ref{ala1}) 
  for $\mu^2 > 0$, one can completely
 fix the renormalization constants.
  Since  ultra-violet (UV) divergences do not dependent on $T$
 in the naive loop-expansion \cite{prf}, 
  $A, B$, $C$ and $D$ are independent of $T$,
 and are expanded as
\beq
\label{counterterms}
 \left(
 \begin{array}{c}  
 A \\ B \\ C \\ D
 \end{array} \right)
  = \sum_{l=1}^{\infty} 
 \left(
 \begin{array}{c}  
 a_l \\ b_l \\ c_l \\ d_l
 \end{array} \right)
 \delta^l.
\eeq
 The coefficients ($a_l, b_l, c_l, d_l$)
 are independent of 
 $\mu^2$, since we use the mass independent renormalization
 scheme.
 Also, the UV divergences in
 the symmetry broken phase $(\mu^2 < 0)$
 can be   removed by the same counter terms determined
 for $\mu^2 > 0$ \cite{BW,kugo}.

 The relations of $A, B, C$ and $D$ with the 
 standard renormalization constants are
  $A = Z-1$, $B= Z_{\mu}Z-1$ and $C = Z_{\lambda}Z^2 -1 $, where
 $Z$'s are defined by  
 $\phi_0 = \sqrt{Z} \phi$, $\lambda_0 =  Z_{\lambda} \lambda$ 
 and $\mu_0^2 = Z_{\mu} \mu^2$ with suffix $0$ indicating 
 unrenormalized quantities.

\noindent
{\bf Step 2}: \\
 Rewrite the Lagrangian (\ref{ala1})  by introducing
 a new mass parameter $m^2$ following the idea of OPT \cite{OPT}:  
\beq
\label{decomp}
 \mu^2 = m^2 - (m^2-\mu^2) \equiv m^2 - \chi. 
\eeq
 This identity should be used  not only in the
 standard mass term but also in the counter terms \cite{point},
 which is crucial to show the order by order renormalization
 in OPT:
\beq
\label{toy2}
  {\cal L}(\phi; \mu^2)   
           & = & {\cal L}_{\rm r} + {\cal L}_{\rm c} \\
\label{toy20}
  {\cal L}_{\rm r}
           & = & {1 \over 2} [(\d \phi)^2 - m^2 \phi^2]
		-{ \lambda \over 4! }\phi^4 + {1 \over 2}\chi \phi^2 \\
\label{toy2c}
 {\cal L}_{\rm c} & = &
     {1 \over 2}A(\d \phi)^2
    -{1 \over 2} B (m^2 - \chi) \phi^2  -{ \lambda \over 4! }C  \phi^4 
   \nonumber \\
 & & + D (m^2-\chi)^2 .
\eeq
 $A$, $B$, $C$ and $D$ in  ${\cal L}_{\rm c}$ were already
 determined in Step 1.

 On the basis of eq.(\ref{toy2}), 
 we define a ``modified'' loop-expansion in which 
 the tree-level propagator has a mass 
 $m^2 + \lambda \varphi^2 /2 $ instead of
 $\mu^2 + \lambda \varphi^2 /2 $.
 Major difference between this expansion and the 
 naive one is the following assignment  
\beq
\label{assign}
 m^2 = O(\delta^0), \ \ \ \  \chi = O(\delta) .
\eeq
 The physical reason behind this assignment  
 is the fact that  $\chi$ reflects the effect of interactions.
 If one adopts an assignment, 
 $ m^2 = O(\delta^0), \chi=O(\delta^0)$,
 the modified  loop-expansion immediately reduces to the naive one.
 
 As will be shown explicitly in  Sec. II.D, all the UV divergences
 in the modified loop-expansion are removed by
 the counter terms determined in the naive loop-expansion.

 Since (\ref{toy2}) 
 is simply a reorganization of the Lagrangian, any Green's functions
 (or its generating functional)  calculated in the 
 modified loop-expansion  should not depend on the arbitrary mass $m$
 if they are calculated in all orders.
 However, 
 one needs to truncate  perturbation series at certain order
 in practice. This 
 inevitably introduces explicit $m$ dependence in  Green's functions.
 Procedures to determine $m$ are given in Step 3 below. 

 To find the ground state of the system,
 one should look for the stationary point of 
 the thermal effective potential $V(\varphi^2)$
 defined by
\beq
 V(\varphi^2) 
= - {\Gamma[\varphi^2 = {\rm const.}]  \over  \int_0^{1/T} d^4 x }.
\eeq
 As mentioned above, $V$ calculated up to 
 $L$-th loops  $V_L(\varphi^2 ;m)$  has explicit $m$-dependence.
 Thus the stationary condition  reads
\beq
\label{del}
  {\partial V_L(\varphi^2;m) \over \partial \varphi } =0 ,
\eeq
 where derivative with respect to  $\varphi$ does not act on $m$
 by definition. Eq.(\ref{del})
 gives a stationary point of $V_L$  for given $m$.

 One may generalize Step 2 by adding and subtracting
 $\alpha_0 (\d_0 \phi)^2$,
 $\alpha_1 (\d_i \phi)^2$
 and $\gamma \phi^4$ \cite{covari}
 with $\alpha_0$, $\alpha_1$ and $\gamma$ 
 being  finite parameters to be determined by the
 PMS or FAC conditions (see Step 3).  $\alpha_0$ and $\alpha_1$ 
 are especially important  for theories with 
 fermions at finite $T$ and chemical potential \cite{saito}.
 We will, however,
 concentrate on the simplest version ($\alpha_{0,1} =\gamma =0$)
 in the following discussions.

\noindent
{\bf Step 3}: \\
 The final step is to find an optimal value of $m$ 
 by imposing physical  conditions  \`{a} la Stevenson \cite{PMS}
 such as the following.
\begin{itemize}
\item[(a)] The principle of 
 minimal sensitivity (PMS):  this condition requires 
 that a chosen
  quantity calculated up to  $L$-th loops  (${\cal O}_L$) should
 be stationary by the variation of  $m$: 
\beq
\label{pmsc}
 {\partial {\cal O}_L \over \partial m } = 0 .
\eeq
 \item[(b)] The criterion of the  fastest apparent convergence (FAC): 
 this condition requires that the perturbative corrections in
 $O_L$ should be as small as possible
 for a suitable value of $m$.
\beq
\label{facc}
 {\cal O}_L - {\cal O}_{L-n} = 0,
\eeq
 where $n$ is chosen in the range, $1 \le n \le L $.
\end{itemize}

 The above conditions
 reduce to self-consistent gap equations whose 
 solution determine
 the optimal parameter $m$ for given $L$. Thus
 $m$ becomes a non-trivial function of $\varphi$, $\lambda$ and $T$ \cite{mt}.
 This together with the solution of (\ref{del}) completely
 determine the thermal expectation value
 $\xi(T) \equiv \langle \phi \rangle$ as well as 
 the optimal parameter $m(T)$. 
 Through this self-consistent process, higher order terms in the
 naive loop expansion  are resumed.

 The choice of ${\cal O}_L$ in Step 3 depends on the quantity one needs
 to improve most.
 To study the static nature of the 
 phase transition, the thermal effective potential
 $V_L(\varphi^2;m)$ is most relevant.
 Applying the PMS condition for $V_L$ reads
\beq 
\label{okopms}
  {\partial V_L(\varphi^2;m) \over \partial m } =0,
\eeq
which gives a solution $m=m(\varphi)$. This can be used to
 improve the effective
 potential at finite $T$ \cite{oko};
\beq
  V_L(\varphi^2; m) \rightarrow
  V_L(\varphi^2; m(\varphi)).
\eeq
 Also, $\xi(T)$ and $m(T)$ are obtained by
 solving (\ref{del}) together with (\ref{okopms}).
  In this case, the following relation holds: 
 $  {d V(\varphi^2;m(\varphi))/ d \varphi }|_{\varphi = \xi} =
 {\d V(\varphi^2;m(\varphi)) / \d \varphi }|_{\varphi = \xi}$.

 To improve particle properties at finite $T$,
 it is  more efficient to
 apply PMS or FAC conditions directly to the two-point functions 
 \cite{coms}.
 In ref.\cite{BM}, FAC with $L=n=2$ was used for
 the boson self-energy calculated up to two-loops.
 We will adopt a similar condition in Sec.III when
 we analyze  spectral functions of the soft modes.

\subsection{UV divergence in the resumed perturbation}

 We briefly mention here  the reason why the renormalization 
 in resumed perturbation is not a trivial issue.

 In the naive perturbation theory,
 there arises no new UV divergences at $T\neq 0$
 because of the natural cutoff from the Boltzmann distribution
 function. Therefore, all the UV divergences at finite $T$
 are canceled by the counter terms prepared at  $T=0$. 
 This statement has been proved in imaginary-time and
 real-time formalisms \cite{prf}.

 On the other hand, in  self-consistent methods at $T \neq 0$,
 the situation is not so simple since
 the tree-level propagators have $T$-dependent  mass (such as 
 $m(T)$ in the above) which contains higher loop contributions
 through the self-consistent gap-equation  \cite{renorm}.
 
 In fact,  in  most of the self-consistent
 methods applied so far (except for ref.\cite{BM}),
 the renormalization is taken into account ``after'' imposing the 
 gap-equation.
 This procedure not only makes the 
 renormalization non-trivial and hard
 in higher orders, but also
 obscures the origin of the UV divergences.
 On the contrary, in OPT explained in the previous subsection,
 the renormalization is performed ``before''
 imposing the gap-equation. In other words,
 the UV divergences are already removed in Step 2, and
 a ``finite'' gap-equation is obtained from the outset in Step 3.  
 
\subsection{Renormalization in  OPT }

 We now prove the order by order renormalization in OPT. 
 Let us first rewrite eq.(\ref{toy2}) as
\beq
\label{ala2}
  {\cal L}(\phi;\mu^2) & = & {\cal L}(\phi;m^2)  \nonumber \\
  &  &  + {1 \over 2} \chi \phi^2 +\left[ {1 \over 2} B\chi \phi^2
  + D \chi^2 - 2D m^2 \chi \right] .
\eeq
 Since we use the symmetric and mass independent
 renormalization scheme (such as the $\overline{MS}$ scheme), 
  any Green's function generated by 
 ${\cal L}(\phi; m^2)$ can be  renormalized
 solely by  the coefficients $A$, $B$, $C$ and $D$
 in ${\cal L}(\phi; \mu^2)$.

 Suppose we make a multiple insertion of  the composite operator
 $(1/2) \chi \phi^2$  to the 
 Green's function generated by  ${\cal L}(\phi; m^2)$.
 The question is whether new divergences
 induced by  the operator insertion
  are made finite  only by the last three counter terms in (\ref{ala2}).
  (Note that $B$ and $D$  are already fixed
 in Step 1, and we do not have any freedom to change them.)

\begin{figure}[h]
%  \begin{center}
\centerline{
    \epsfxsize=7.4cm
    \epsfbox{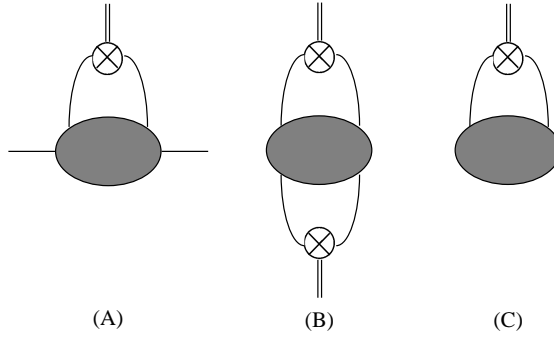}
}
%  \end{center}
     \caption{Diagrams which contain UV divergences as a result
     of the multiple insertion of $(1/2)\chi \phi^2$.
    (A) corresponds to a single insertion with two external lines.
    (B) and (C) have no external lines with
     a single insertion and a double insertion, respectively.}
 \label{comr}
\end{figure}

 The above problem is obviously related to the 
 renormalization of composite operators.
  In fact,  the standard method \cite{IZ} tells us that
 necessary counter terms 
 to remove the divergences induced by the insertion of $(1/2)\chi \phi^2$
 are written as 
\beq
\label{count2}
 {1 \over 2 } (Z Z^{-1}_{\phi^2} -1) \chi \phi^2 + \Delta_2 \chi^2 
  + \Delta_1 \chi . 
\eeq
 Here $Z_{\phi^2}$ is the renormalization constant for the
 composite operator $\phi^2$, and is necessary to remove the 
 divergence in Fig. \ref{comr}(A).
 $\Delta_2$ and $\Delta_1$ are necessary 
 to remove the overall divergences
 in Fig. \ref{comr}(B) and in Fig. \ref{comr}(C), respectively.

 Now, one can prove that (\ref{count2}) coincides with 
  the last three terms in (\ref{ala2}):
\beq
\label{eqq}
  Z Z^{-1}_{\phi^2} -1 = B, \ \ \ \Delta_2 = D, \ \ \ {\rm and } \ \ \ 
  \Delta_1 = -2Dm^2.
\eeq
 The first equation is obtained by the 
 definition $B= Z_{\mu}Z-1$ and an identity 
\beq
\label{zet1}
 Z_{\phi^2} = Z_{\mu}^{-1}.
\eeq
 The overall divergence of the vacuum diagram with
 no external-legs is removed by the $c$-number counter
 term $Dm^4$ in ${\cal L}(\phi; m^2)$. Therefore,  
 the last two equations in (\ref{eqq}) are obtained as
\beq
\label{zet2}
 \Delta_1 & =  &
  - ({\partial \over \partial m^2}) \left[ Dm^4 \right] = -2D m^2 ,\\ 
\label{zet3}
 2 \Delta_2 & = &
  ({\partial \over \partial m^2})^2 \left[ Dm^4 \right] = 2D.
\eeq
 For completeness, an explicit proof
 of (\ref{zet1},\ref{zet2},\ref{zet3}) is given in
  APPENDIX A.

  Eq.(\ref{eqq}) shows clearly that all the
  necessary counter terms in OPT
 are supplied solely by the original Lagrangian ${\cal L}(\phi; \mu^2)$.  
 Let us now define $\chi^j \Gamma^{(n,j)}_R (\lambda, m^2)$ as 
 a renormalized $n$-point proper vertex with insertion
 of  $(1/2) \chi \phi^2$ by $j$-times.
 (Here the external momentums are not written explicitly.)
 The counter terms in (\ref{count2}) together 
 with (\ref{eqq}) assure the finiteness of 
 $\Gamma^{(n,j)}_R$.  Since the proper vertex can be expanded
 as $\chi^j \Gamma^{(n,j)}_R = \delta^j \sum_{l=0}^{\infty} \gamma_{l}
\delta^{l}$, each coefficient $\gamma_l$ is also finite.
 This implies that
  $\chi^j \Gamma^{(n,j)}_R$ can be made finite order by order in OPT.

\vspace{0.2cm}

Three comments are in order here:
\begin{itemize}
\item[(i)] Because the renormalization is already carried out
 in Step 2, one obtains finite gap-equations from the beginning
 in Step 3.  Our procedure ``resummation after renormalization''
 has several advantages over the conventional procedure 
``resummation before renormalization''
  where UV divergences are hoped to be canceled
 after imposing the gap-equation.
 The difference between the two is prominent 
 especially in higher order calculations. 
\item[(ii)]
 The decomposition (\ref{decomp}) should be done both
 in  the mass term and the counter terms.
  This guarantees the order by order
 renormalization in our modified loop-expansion.
 In ref.\cite{BM},  the order by order renormalization
 was checked up to the two-loop order in the $\phi^4$ theory
 at high $T$.  Our proof shows that this nice feature
 holds in any higher orders in OPT.
  On the other hand, if one keeps the
 original counter term $(1/2) B \mu^2 \phi^2 + D \mu^4$
 without the decomposition (\ref{decomp}),
  $L$-loop diagrams with $L > M$ must be taken into account 
 to remove the UV 
 divergences in the $M$-loop order
  (see e.g. the last reference in \cite{BP}). 
 This is an unnecessary complication due to the 
 inappropriate treatment of the counter terms.
\item[(iii)]
 As far as we stay in the low energy region far below the
 Landau pole, we need not address the issue of
 the triviality of the $\phi^4$ theory \cite{triv}: Perturbative
 renormalization in OPT  works
 in the same sense as that in the naive perturbation.

\end{itemize}

\subsection{Nambu-Goldstone theorem}

 The procedure and the renormalization in OPT  discussed above
 do not receive modifications even if
 the Lagrangian has global symmetry.
 For $O(N)$ $\phi^4$ theory, one needs to replace
 $\phi$  and $\phi^2$
 by $\vec{\phi}
 = (\phi_0, \phi_1, \cdot \cdot \cdot , \phi_{N-1})$
 and $\vec{\phi}^2$ respectively 
 in all the previous formulas.
 In the symmetry broken phase of such theory,
 the Nambu-Goldstone (NG) theorem and massless
 NG bosons are guaranteed in each order of the modified 
 loop-expansion in OPT for arbitrary $N$. 
  To show this, it is most convenient to start
 with the thermal effective potential $V(\vec{\varphi}^2)$.
  By the definition of the 
 effective potential,  $V(\vec{\varphi}^2)$ has
 manifest $O(N)$ invariance if it is calculated in all orders.
 
In OPT,  $V$ calculated up to
 $L$-th loops $V_L(\vec{\varphi}^2;m)$
 has also manifest $O(N)$ invariance, because
 our decomposition (\ref{decomp}) used in  (\ref{toy2})
 does not break  $O(N)$ invariance.
 Once $V_L$ has 
 invariance under the $O(N)$ rotation
 ($\varphi_i \rightarrow \varphi_i + i \theta^a T^a_{ij} \varphi_j$),
 the immediate consequence is the standard identity: 
\beq
\label{NG1}
 {\d V_L(\vec{\varphi}^2;m) \over \d \varphi_j } T_{ji}^a
 = - {\d^2 V_L(\vec{\varphi}^2;m) \over \d \varphi_i \d \varphi_j } 
 T_{jk}^a \varphi_k,
\eeq
 with ${\bf T}^a $ being the generator of the $O(N)$ symmetry.
 Eq.(\ref{NG1})  is valid for arbitrary $L$, $m$ and $N$.

 At the stationary point where the l.h.s. of (\ref{NG1})
 vanishes, there arises  massless NG bosons
 for $T_{jk}^a \varphi_k \neq 0$, since
 the r.h.s. of  (\ref{NG1}) is equal to  
 $ -{\cal D}_{ij}^{-1}(0) T_{jk}^a \phi_k $ where
 ${\cal D}_{ij}(0)$ is the Matsubara 
 propagator at zero frequency and momentum calculated
 up to $L$-th loops.  Thus the existence of the
 NG bosons is proved independent of the structure
 of the gap-equation in Step 3.

  It is instructive here to show some unjustified approximations
  which lead to the breakdown of the NG theorem.
   Many of the self-consistent methods applied so far
 suffer from these problems \cite{okog}.
 
\begin{enumerate}
\item[(i)] 
 Suppose one takes into account only a part of the diagrams 
 for given number of loops. Then the  pion is no longer massless
 even if the symmetry is spontaneously broken.
 Although this is a trivial point, sometimes
 such approximation is adopted in the literatures:
 taking only the self-energy from the 
 four-point vertex and neglecting that from the three-point  
 vertex is a typical example.
\item[(ii)]
 Introducing  $m^2$ in the $O(N)$ symmetric way
 as (\ref{decomp})
  is a key for the NG theorem to hold in each order of the 
 loop-expansion in OPT.  Suppose that
 we make a general decomposition such as 
\beq
\label{abs}
  \mu^2 \delta_{ij} = m^2_{ij} - ( m^2_{ij} - \mu^2 \delta_{ij}),
\eeq
 with $m^2_{ij} \neq m^2 \delta_{ij}$.
 This leads to 
 an $O(N)$ non-invariant effective potential, 
 and the relation (\ref{NG1}) is not guaranteed
 in any finite orders of the loop-expansion.
 For example, when the $O(N)$ symmetry is
 spontaneously broken down to $O(N-1)$,
 one may be tempted to make a decomposition
 $m_{ij}^2 = m_0^2 $ ($i$=$j$=0), $m_{ij}^2 = m_1^2 $ ($i$=$j$$\neq$0),
 and $m_{ij}^2 = 0 $ ($i$$\neq$$j$).
 This leads to an effective potential
 $V_L(\varphi_0^2, \sum_{i=1}^{N-1}\varphi_i^2;m_0,m_1)$
 which has only $O(N-1)$ invariance.
 It implies that eq.(\ref{NG1}) hold only for generators
 which do not mix $\varphi_0$ with $\varphi_{i\neq 0}$.
 However, eq.(\ref{NG1}) for those generators alone
 are not enough to prove
 the existence of NG bosons: In fact, $\varphi_{k\neq 0}$
 in the r.h.s. of (\ref{NG1}) vanishes 
 in the $O(N-1)$ symmetric ground state,
 and no constraints are obtained for  ${\cal D}_{ij}^{-1}(0)$.
\end{enumerate}

\section{Application to $O(4)$ $\sigma$ model}

 Let us apply OPT to study the soft mode associated with 
 chiral transition.  Our main goal is to investigate the 
 spectral functions of the soft modes at finite $T$: 
\beq
\label{spectral}
   \rho_{\phi}(\omega, {\bf k};T)  = 
   - {1 \over \pi} \mbox{Im} D_{\phi}^{R}(\omega, {\bf k};T) .
\eeq
 Here $D_{\phi}^{R}$ is the retarded
 correlation function
\beq
  D_{\phi}^{R}(\omega, {\bf k};T) 
    =  -i  \int d^4x e^{i kx}
   \theta(t) \la  [\phi(t,{\bf x}), \phi(0,{\bf 0})] \ra,
\eeq  
 where  $\la \cdot \ra$ denotes thermal expectation value, and
  $ \phi(t,{\bf x})$ is $\bar{q}q (t,{\bf x})$ or
 $\bar{q}i \gamma_5 \vec{\tau } q (t,{\bf x})$ in QCD. 

 This spectral function was first studied 
 using the Nambu-Jona-Lasinio model of QCD in the
 large $N_c$ limit \cite{HK85}.
 The analysis shows that the scalar meson $\sigma$, which
 has a large width due to the strong decay
 $\sigma \rightarrow 2 \pi$,
 decreases its mass as $T$ increases. Eventually, $\sigma$   
 shows up as a sharp resonance near the critical point of 
  chiral transition.
 The detectability of such resonance
 was studied in the context of 
 the  ultra-relativistic heavy ion collisions
 \cite{welk}. Also, the spectral integrals in QCD at finite $T$
 were studied using the operator product expansion \cite{huang}.

 In the following, we will adopt a toy model 
 ($O(4)$ linear $\sigma$ model) to study the
 effect of mode couplings (interaction among the soft modes)
 in the one-loop level at finite $T$.
 This model shares common symmetry and dynamics
 with QCD and has been used to study the 
 real-time dynamics and critical phenomena \cite{CH,PW}.

\subsection{Determination of the parameters at $T=0$}

 The $O(4)$ $\sigma$ model reads
\beq
\label{lin1}
  {\cal L} & = & 
  {1 \over 2} [(\d \vec{\phi})^2 - \mu^2 \vec{\phi}^2]
  -{ \lambda \over 4! }(\vec{\phi}^2)^2 + h \sigma  \nonumber \\
& &\mbox{}+{1 \over 2} A(\d \vec{\phi})^2 -{1 \over 2} B \mu^2 \vec{\phi}^2 
  -{ \lambda \over 4! }C(\vec{\phi}^2)^2+D \mu^4 ,
\eeq
 with $\vec{\phi}=(\sigma, \vpi)$.
 $h \sigma$ is an explicit chiral symmetry breaking term \cite{hs}.
%
% on account of non-zero quark mass and the pion mass is controlled by 
% this term.
%
 $A$, $B$, $C$ and $D$ in the one-loop order are
\beq
\label{counter}
  A =0, \ \ \ 
  B = {\lambda \over 16\pi^2} \bep , 
  \ \ \ C={\lambda \over 8\pi^2}\bep , 
  \ \ \  D = -{1 \over 16 \pi^2} \bep ,
\eeq
 where $ \bep \equiv {2 \over 4-n}-\gamma+\log(4\pi)$,
 with $\gamma$ being the Euler constant.

 When SSB takes place $(\mu^2 < 0)$,
 the replacement $\sigma \rightarrow \sigma + \xi$
 in  eq.(\ref{lin1}) leads to the tree-level masses of
 $\sigma $ and $\pi$;
\beq
\ms ^2= \mu^2+\frac{\lambda}{2}\xi^2, \ \ 
\mp ^2= \mu^2+\frac{\lambda}{6}\xi^2 .
\eeq
 The expectation value $\xi$ at $T=0$ is determined by the stationary 
 condition for the standard effective potential 
 $\d V(\vec{\varphi})/\d \varphi_j =0 $.

 Later we will take a special FAC condition in which 
 $m^2$ deviates from $ \mu^2$ only at $T \neq 0$, so that
 the naive loop-expansion  at $T=0$ is valid.
 The renormalized couplings $\mu^2, \lambda$ and $h$
 can thus be determined by the following physical conditions 
 in the naive loop-expansion at zero $T$:
\begin{enumerate}
\item[(i)] On-shell condition for the pion
   $D_{\pi}^{-1}(k^2=m_{\pi}^2)=0$, where 
   $m_{\pi}=140 $ MeV, and $D_{\pi}$ is the 
   causal propagator for the pion in one-loop order.
\item[(ii)] Partially conserved axial-vector current (PCAC) 
   relation in one-loop; $f_{\pi}m_{\pi}^2=h \sqrt{Z_{\pi}}$. Here
   $f_{\pi}=93$ MeV, and  $Z_{\pi}$ is the {\em finite}
   wave function renormalization constant for the pion on its mass-shell. 
\item[(iii)] The peak position of the spectral function
   in the $\sigma$ channel ($ \equiv m^{peak}_{\sigma}$)
   is  taken to be 550 MeV, 750 MeV or 1000 MeV.
\end{enumerate}

 $m^{peak}_{\sigma}=$ 550 MeV in (iii)
 is consistent with recent re-analyses of the 
 $\pi$-$\pi$ scattering phase shift \cite{pipi}. However,
 our main conclusions
 do not suffer qualitative change by other choices, 750 MeV and 1000 MeV.
 Instead of $m_{\sigma}^{peak}$, one may
  take the $\pi$-$\pi$ scattering phase shift itself
 as a condition to determine parameters  \cite{CH}.
 However, for the discussions in the following,
 such sophistication is not necessary. 
 
 We still have a freedom to choose the 
 renormalization point $\kappa$.
 Instead of trying to determine optimal $\kappa$
 by the renormalization group equation for the
 effective potential \cite{kast}, we take a simple and 
 physical condition
 $m_{0\pi}$=$m_{\pi}$=140 MeV which is 
 suitable for our later purpose. 
 This choice has two advantages: (a) The one-loop
 pion self-energy $\Sigma_{\pi}(k^2)$ vanishes
 at the tree-mass; $ \Sigma_{\pi}(k^2 = m_{0 \pi}^2) 
 = \Sigma_{\pi}(k^2=m_{\pi}^2)=0$, where we have used the 
 condition (i) together with $m_{0\pi}=m_{\pi}$.
 (b) The spectral function in the $\sigma$ channel
 starts from a correct continuum threshold 
 in the one-loop level. (In the loop-expansion,
 $\rho_{\sigma}(\omega,{\bf 0})$ has a physical 
 threshold at $\omega = 2
 m_{\pi}$=280 MeV only if $m_{0\pi} = m_{\pi}$.)

\begin{figure}[h]
%  \begin{center}
\centerline{
    \epsfysize=6.3cm
    \epsfbox{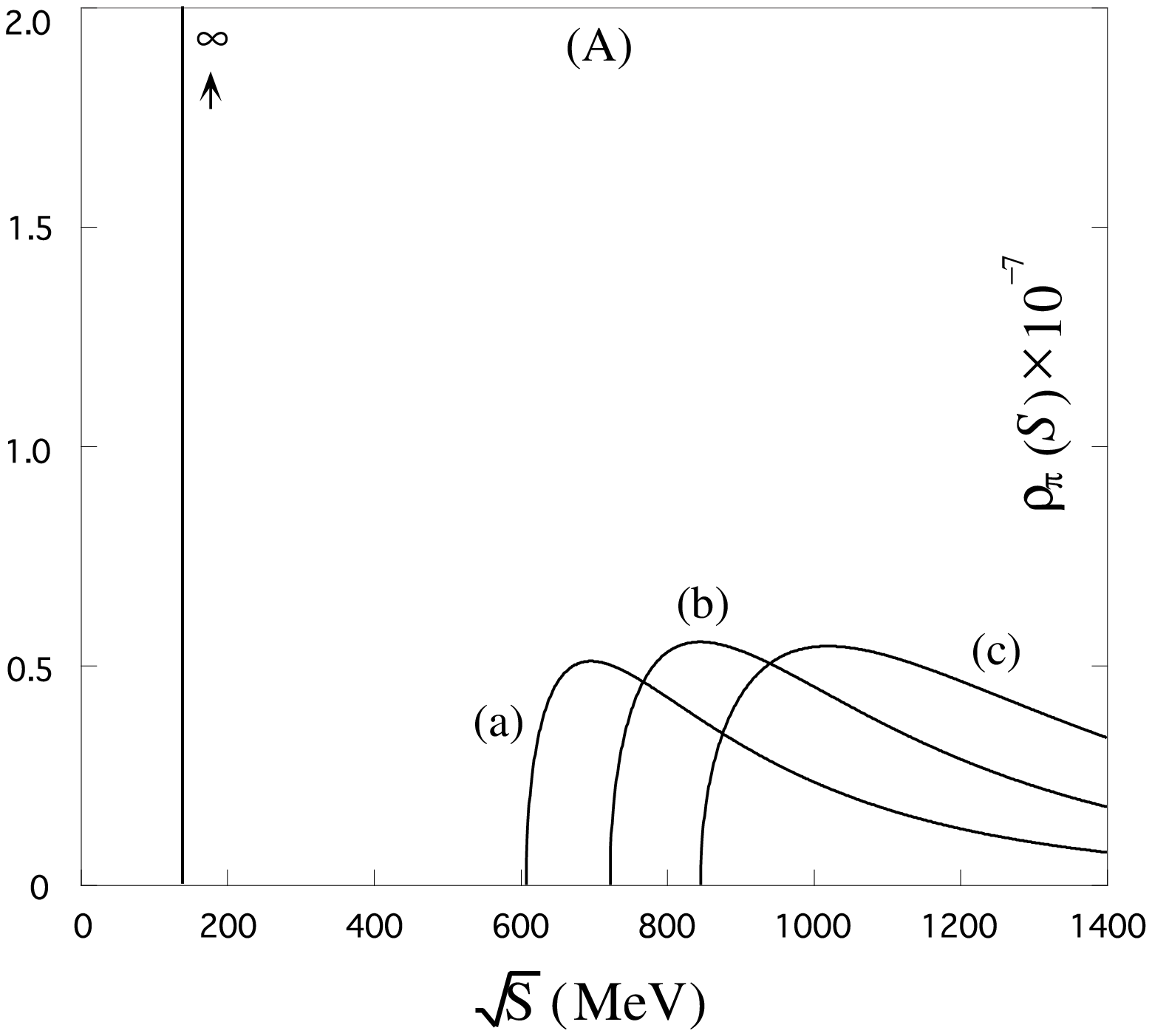} \hspace{0.2cm}
    \epsfysize=6.3cm
    \epsfbox{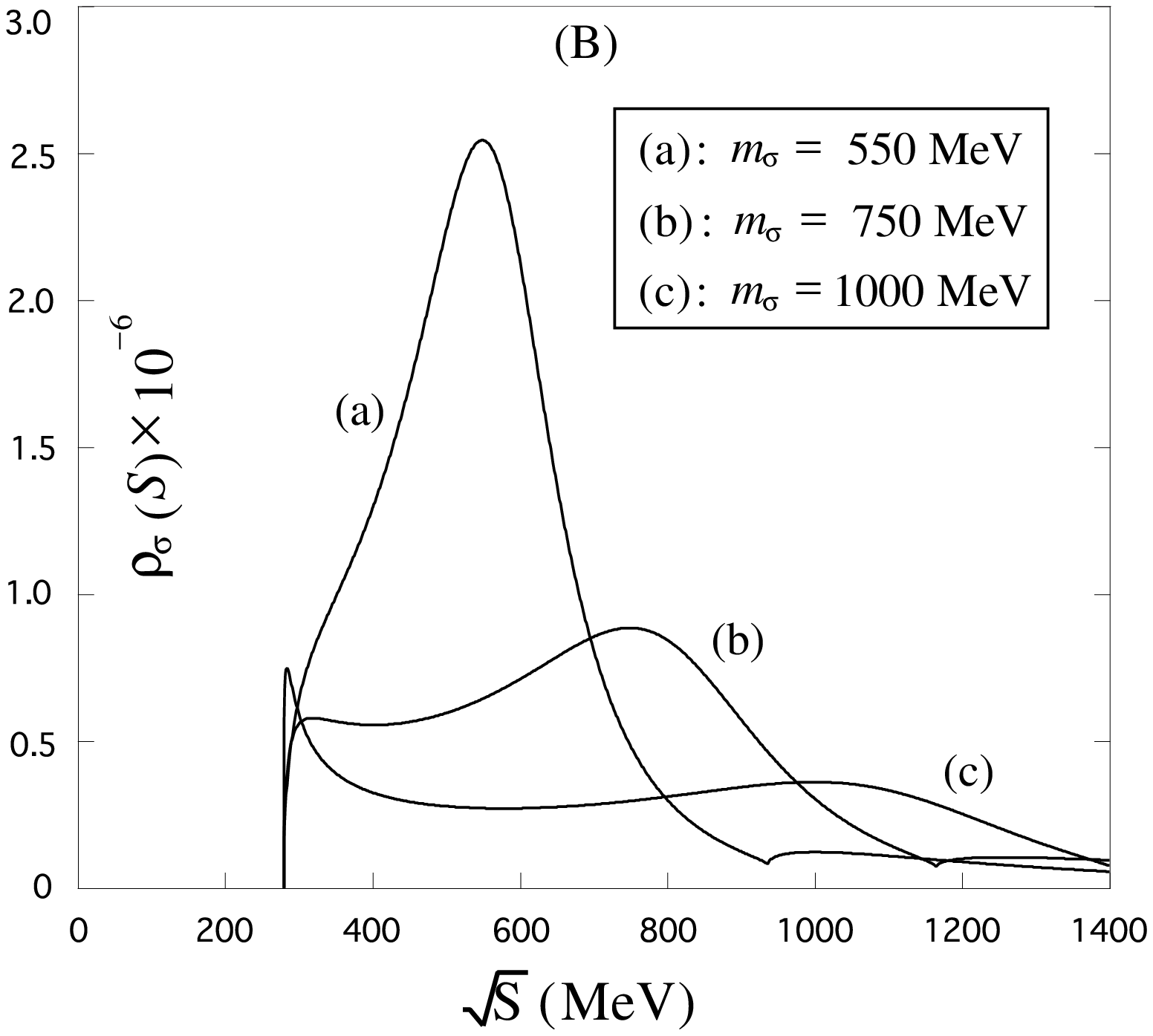}
}
%  \end{center}
     \caption{Spectral functions at $T=0$ in the 
     $\pi$ channel (A) and in the $\sigma$ channel
     (B) for $m^{peak}_{\sigma}=550$ MeV, 750 MeV and  1000 MeV.}
 \label{spect}
\end{figure}

 Resultant  parameters are summarized  in TABLE I.
 The spectral functions $\rho_{\sigma}$ and $\rho_{\pi}$ defined in
 (\ref{spectral}) at $T=0$ with $s \equiv \omega^2 - {\bf k}^2$
 are shown in Fig. \ref{spect}.
 In the $\pi$ channel, there are one particle pole and 
 a continuum  starting from the threshold $\sqrt{s_{th}}
 = m_{0 \pi} + m_{0 \sigma}$. $\sqrt{s_{th}}$ 
 is the point where the channel $\pi + \sigma$ opens.
 In the $\sigma$ channel, the spectral function starts
 from the threshold $2 m_{0 \pi} = 280$ MeV and shows 
 a broad peak centered around $\sqrt{s} = m^{peak}_{\sigma}$.
 The half width of the peak is 260 MeV, 657 MeV and 
 995 MeV for $m_{\sigma}^{peak} =$ 550 MeV, 750 MeV and 1000 MeV,
 respectively. 
 The large width of $\sigma$ 
 is due to a strong $\sigma$-$2\pi$ coupling  
 in the linear $\sigma$ model. The corresponding $\sigma$-pole
 is located far from the real axis on the complex $s$ plane.

\subsection{Application of OPT}

 Now let us proceed to
 Step 2 in OPT and rewrite eq.(\ref{lin1})
 as 
\beq
\label{lag2}
  {\cal L} & = & {1 \over 2} [(\d \vec{\phi} )^2 -m^2 \vec{\phi}^2]
   -{ \lambda \over 4! } (\vec{\phi}^2)^2 
   + {1 \over 2} \chi \vec{\phi}^2 
  + h \sigma \nonumber\\
  & &  - {1 \over 2} B m^2 \vec{\phi} ^2 
 -{ \lambda \over 4! } C (\vec{\phi}^2)^2 
       + D m^4.
\eeq
 Since $\chi$ ( = $m^2-\mu^2$) is already a one-loop order,
 we have neglected the terms 
 proportional to $B \chi$, $D \chi^2$ and $D \chi$
 which are two-loop or higher orders.
%
% Renormalization is trivial in this order because the new vertex term
% $\chi$ does not make divergent.
%

 When SSB takes place ($\sigma \rightarrow \sigma + \xi$),
 the tree-level masses to be used in the modified loop-expansion read
\beq
\label{Rbmass}
  \ms ^2=m^2+\frac{\lambda}{2}\xi^2, \ \  
  \mp ^2=m^2+\frac{\lambda}{6}\xi^2.
\eeq
 Since $m^2$ will eventually be a function of $T$,
 the tree-masses running in the loops
 are not necessary tachyonic at finite $T$ contrary to the 
 naive loop-expansion (see the discussion in Sec. II.A).

 The thermal effective potential $V(\vec{\varphi};m)$
 is calculated in the standard manner except for
 the extra terms proportional to $\chi$.
 The Gibbs free energy $G(\xi;m) \equiv 
 V(\vec{\varphi} = (\xi,{\bf 0});m) $ in the one-loop level 
 reads
\beq
\label{pot}
&  & G(\xi;m)  = \nonumber \\
& \ \  & {1 \over 2} \mu^2 \xi^2 + {\lambda \over 4!}\xi^4
   - h \xi 
\nonumber \\
 &\ \  & + {1 \over 64 \pi^2} \left[ 
 m_{0 \sigma}^4 \  
 {\rm ln} \left| { m_{0 \sigma}^2 \over \kappa^2 e^{3/2} } \right|
+ 3 m_{0 \pi}^4 \ 
 {\rm ln} \left| { m_{0 \pi}^2 \over \kappa^2 e^{3/2}} \right|   
 \right]
 \\
 & \ \ & + T \int {d^3k \over (2 \pi)^3} \left[
 {\rm ln} 
  (1- e^{- E_{\sigma} /T} ) 
 + 3\  {\rm ln} 
   (1- e^{- E_{\pi} /T} ) \right] \nonumber,
\eeq
 where $E_{\phi} \equiv \sqrt{ {\bf k}^2 + m_{0 \phi}^2 }$.
 Although this has the similar structure with the standard free energy
 in the naive loop-expansion,
 the coefficient of the first term in the r.h.s.
 of (\ref{pot})  is
 $\mu^2$ instead of $m^2$. This is because
 we have extra mass-term proportional to
 $\chi$ in the one-loop level.  The stationary point $\xi$  
 is obtained by 
\beq
\label{dvev}
 \left. {\d V(\vec{\varphi} ;m) \over \d \varphi_i} \right| _{\vec{\varphi} 
 = (\xi, {\bf 0})} = {\d G(\xi;m) \over \d \xi } =0.
\eeq
 Since the derivative with respect to $\xi$ does not act on $m$,
 this gives a solution $\xi$ as a function of $T$ and $m$.
 By imposing another condition on $m$ (Step 3), one eventually
 determines both $\xi$ and $m$ for given $T$.

 At finite $T$, the retarded propagator has a general form
\beq
\label{proprr}
  iD^R_{\phi}(\omega,{\bf k};T) = {i \over k^2 - m_{0\phi}^2 - 
  \Sigma^R_{\phi} (\omega, {\bf k};T)},
\eeq
 with  $k^2 = \omega^2 - {\bf k}^2$.
 The spectral function is then written as
\beq
\label{spect2}
  \rho_{\phi}(\omega,{\bf k};T)
   =  - {1 \over \pi}\ {{\rm Im} \Sigma^R_{\phi} \over
  (k^2 - m_{0\phi}^2 - {\rm Re}\Sigma^R_{\phi})^2 +
  ({\rm Im}\Sigma^R_{\phi} )^2 }.
\eeq

 The retarded self-energy $\Sigma^R_{\phi}$ is related to the
 11-component of the 
 2 $\times$ 2 self-energy in the real-time formalism \cite{real};
\beq
  \label{R-11}
  {\rm Re} \Sigma^R_{\phi} (\omega, {\bf k};T)& = &
  {\rm Re} \{\Sigma^{11}_{\phi}(\omega,{\bf k})+
  \Sigma^{11}_{\phi}(\omega, {\bf k};T)\}  \\ 
  {\rm Im} \Sigma^R_{\phi} (\omega, {\bf k};T)& = &
  \tanh({\omega \over 2T})\  {\rm Im} \{\Sigma^{11}_{\phi}(\omega,{\bf k})+
  \Sigma^{11}_{\phi} (\omega, {\bf k};T)\}. \nonumber
\eeq
 Here  $\Sigma^{11}_{\phi}(\omega,{\bf k};T)$
 is defined as a part with explicit
$T$-dependence
 through the Bose-Einstein distribution, while
 $\Sigma^{11}_{\phi}
 (\omega, {\bf k})$ is the part which has only implicit $T$-dependence
 through $m(T)$ and $\xi(T)$.
  In the one-loop level,
 $\Sigma^{11}_{\phi}$ can be calculated only by the 11-component of the
 free propagator,
\beq
\label{T-pro}
  iD_{0 \phi}^{11}(k^2;T) =  {i \over k^2-m_{0 \phi}^2+i\epsilon} 
   + 2 \pi n_{B} \delta(k^2-m_{0 \phi}^2),
\eeq
 with $n_{B}=[e^{\omega /T}-1]^{-1}$.

 One-loop diagrams in OPT for  $\Sigma^{11}_{\phi}$
 are shown in Fig. \ref{self}. Their explicit forms
 are given in APPENDIX B.
 The NG theorem discussed in Sec.II.E 
 can be explicitly checked
 by comparing eq.(\ref{dvev}) and the inverse 
 pion-propagator at zero momentum
 $[D^{R}_{\pi}(0,{\bf 0};T)]^{-1} $.

\subsection{Cancellation of $T$-dependent infinities}

 It is instructive here to show explicitly how the 
 UV divergences discussed in Sec.II.D
 are canceled in the one-loop order.
 The divergent part of $\Sigma_{\pi}^R(\omega, {\bf k};T)$
 from diagrams Fig.\ref{self}($h, i, j$)  reads
\beq
& &\Sigma_{\pi}^R [(h)+(i)+(j)]  \nonumber \\
 & \rightarrow &   
 - {\lambda  \over 16 \pi^2 \bar{\vep}} 
 \left( { 5  \over 6} m_{0 \pi}^2(T) +  {1  \over 6} m_{0 \sigma}^2(T) 
 +{\lambda \xi^2(T) \over 9} \right)  \nonumber \\
 & & =  - {\lambda  \over 16 \pi^2 \bar{\vep}} 
 \left( m^2(T) + {1 \over 3} \lambda \xi^2(T) \right) \nonumber \\
 & & = - \Sigma_{\pi}^R [(l)+(m)] ,
\eeq
where eq.(\ref{Rbmass}) has been used.
 Namely,
 the terms proportional to $m^2(T)$ in
 Fig.\ref{self}($h, i)$  are canceled by the counter term
 proportional to $Bm^2$, while the terms proportional to
 $\xi^2(T)$ in Fig.\ref{self}($h, i, j)$ are canceled
 by the usual counter terms proportional to $C$.
 In this way, the $T$-dependent divergences proportional to
 $m^2(T)$ newly appeared in OPT is automatically
 canceled by the $T$-dependent counter terms 
 obtained by the shift $\mu^2 = m^2 - \chi$.
 The divergence of the
 free energy proportional to $m^4$ is also canceled
 by the last counter term in (\ref{ala2}).

 Note that the divergences proportional to
 $\chi$ start to appear
  from the two-loop level. They are removed by the counter terms
 proportional to $\chi$ obtained by the shift
 $\mu^2 = m^2 - \chi$.

\subsection{FAC condition for $m^2$}

 Since we are interested in the spectral functions
 in the one-loop level, a best way 
 to determine $m^2$ is to use the two point function in the
 $\pi$ channel.
 In \cite{BM},
 a FAC condition (\ref{facc}) for 
 the two-loop self-energy at zero momentum ($L=n=2$) was taken to
 obtain a gap equation for
 $\phi^4$ theory above $T_c$.

 The corresponding condition in our model with $L=n=1$ reads
\beq
\label{BMC}
 & &\Sigma_{\pi}^R ( \omega=0, {\bf 0}; T)  \nonumber \\
 & & \ \ \ =  \Sigma_{\pi}^{11} ( \omega=0, {\bf 0}) 
 +  \Sigma_{\pi}^{11} ( \omega=0, {\bf 0}; T) = 0.
\eeq
This is a condition that the one-loop
 correction to the self-energy must be as small as possible 
 in the resumed perturbation theory. 
 ( Note that ${\rm Im} \Sigma_{\pi}^R ( \omega=0, {\bf 0}; T)$ vanishes
 identically.) 
 Unfortunately, (\ref{BMC})  is incompatible with 
 the condition which we adopted at $T=0$ in Sec.III.A to find optimal 
 renormalization point $\kappa$:
\beq
\label{kap}
\Sigma_{\pi}^R ( \omega=m_{0 \pi}, {\bf 0}; T=0) = 0.
\eeq
 A hybrid condition which does not destroy 
 eq.(\ref{kap}) and simultaneously leads to a valid 
 gap-equation is
\beq
\label{CHC}
 \Sigma_{\pi}^{11} ( \omega=m_{0 \pi}, {\bf 0}) 
 +  \Sigma_{\pi}^{11} ( \omega=0, {\bf 0}; T) = 0.
\eeq
The explicit form of this FAC condition can be read 
 from Eq.(B2) in APPENDIX B:
\beq
\label{meq}
m^2 & = & \mu^2 \nonumber \\
 & & +
   {\lambda \over 6} \left[ 5 \tilde{I}_{\pi}^{(1)} + \tilde{I}_{\sigma}^{(1)}
  -i {2 \over 3} \lambda \xi^2 \ \tilde{I}^{(3)} 
   \right] _{ \omega=m_{0 \pi}} \nonumber \\ 
 & & + {\lambda \over 6} \left[ 5 F_{\pi}^{(1)} + F_{\sigma}^{(1)}
  -i {2 \over 3} \lambda \xi^2 \ (F^{(4)} + F^{(5)}) \right]
  _{ \omega=0} .
\eeq
 The second (third) line is from  the first (second) term in 
 the l.h.s. of eq.(\ref{CHC}). The functions 
 $I$ and $F$ are given in APPENDIX B ($\tilde{I}$ is defined as 
 the finite part of $I$.)

 At $T$=0, the second term in the l.h.s. of eq.(\ref{CHC}),
 $\Sigma_{\pi}^{11} ( \omega, {\bf k}; T=0)$, vanishes
  by definition, and eq.(\ref{CHC}) formally reduces to
 eq.(\ref{kap}). However, we calculate
 (\ref{kap}) in the naive loop-expansion 
 without introducing $m^2$ as discussed in Sec.III.A,
 while  (\ref{CHC}) is calculated with $m^2$
 even at $T=0$. Therefore, they are consistent only when
\beq
\label{self1}
  m^2(T=0) = \mu^2.
\eeq 
In other words, OPT with the FAC condition (\ref{CHC})
 applied at $T=0$ is equivalent to the
 naive-loop expansion.
 
 In the symmetric phase at high $T$ where $\xi(T) \simeq 0$,
 eq.(\ref{meq}) reduces to 
\beq
\label{self2}
 m^2 = \mu^2 + \lambda 
 \left[
 \int{d^3 k \over (2 \pi)^3 } {n_B(E(m)) \over E(m)}
  + {m^2 \over 16 \pi^2} \ln {m^2 \over e \kappa^2} 
\right] ,
\eeq
 with $E(m)=\sqrt{m^2 + {\bf k}^2}$.
 If $T^2 \gg m^2$, the first term in the r.h.s. of
 eq.(\ref{self2}) dominates and the following solution is
 obtained:
\beq
\label{hightt}
  m^2 (T) = \mu^2 + {\lambda \over 12} T^2 ,
\eeq
 which implies that the Debye screening mass at high $T$ can be
 properly taken into account. Also, 
 both eq.(\ref{CHC}) and eq.(\ref{BMC}) 
 have the same solution  (\ref{hightt}) for $T^2 \gg m^2$ 
 and are consistent with each other. 
  For realistic values of  $\lambda$ in TABLE I, the condition
 $T^2 \gg m^2$ is not well satisfied and 
 one needs to solve eq.(\ref{CHC}) numerically which will be
 shown in Sec. III.E.
 
 For intermediate values of $T$, eq.(\ref{CHC}) can effectively
 sum not only the contributions from 
 the diagrams in Fig.\ref{self}$(a,b,h,i)$, 
 but also from those in Fig.\ref{self}$(c,d,j)$.
 Thus, OPT can go beyond the cactus
 approximation which sums only Fig.\ref{self}$(a,b,h,i)$.

\vspace{0.2cm}

 Three remarks are in order here.
\begin{enumerate} 
\item[(i)]
 For sufficiently high $T$ with fixed $\kappa$,
 eq.(\ref{self2})
 ceases to have a solution.  In fact,
 the r.h.s. of  eq.(\ref{self2}) is always larger than  the l.h.s.
 above a limiting temperature  $T_l = 500, 430, 420$ MeV 
 for $m_{\sigma}^{peak}(T=0) = 500, 750, 1000$ MeV, respectively.
 This indicates that our one-loop analysis
 in OPT is not sufficient for $T > T_l$. 
 One may try the renormalization group improvement
 by choosing e.g. $\kappa = T$ to cure this problem.
 However, since $\lambda$ in TABLE I  is rather large,
 one encounters the Landau pole in the running coupling
  $\lambda(T)$
 located at $T = 440, 450, 490$ MeV for 
 $m_{\sigma}^{peak}(T=0)= 500, 750, 1000$ MeV, 
 respectively. This again sets an upper bound of $T$
 beyond which the one-loop analysis in OPT is not reliable.
\item[(ii)]
 To study the possible variations of the
 FAC condition  eq.(\ref{CHC}),
 we have examined the following three cases.
 (a) Taking the high $T$ formula eq.(\ref{hightt})
 in place of  eq.(\ref{CHC}).
 (b)  Replacing the second term in  the l.h.s. of (\ref{CHC}) by
 $C_1 \equiv \Sigma_{\pi}^{11} ( \omega=m_{0 \pi}, {\bf 0}; T)$.
 (c)  Replacing the second term in  the l.h.s. of (\ref{CHC}) by
 $C_2 \equiv {\rm Re} \Sigma_{\pi}^{11} (\omega=m_{0 \pi}, {\bf 0}; T)$.
 In the case (a), because of the lack of self-consistency at low $T$,
  $m_{0 \pi}(T)$ deviates substantially 
  from $m_{\pi}(T)$. This leads to an incorrect threshold 
 for the  spectral-function in the $\sigma$ channel.
 In the case (b), 
 a real solution for $m^2$ is not guaranteed, because 
 $C_1$ is a complex function due to the Landau damping.
 In the case (c), a discontinuity of $m^2$ at certain
 $T$ appears, because $C_2$ has a cusp structure as a function of 
 $\omega$ as shown in Fig. \ref{spectT} (C).
  (We will discuss this cusp in Sec.III.G.) 
 Therefore, within the FAC condition for the one-loop self-energy,
 eq.(\ref{CHC}) is almost a unique choice in the sense that 
 it gives a smooth and physically acceptable solution for $m^2$. 
\item[(iii)]
 In place of the FAC condition, one may take the PMS
 condition. However, 
 to get a sensible gap-equation from the PMS condition
 for the thermal effective potential $V$, one needs to calculate
 $V$ at least up to two-loops \cite{oko}.
 Unlike the optimized expansion considered in the first reference
 in \cite{oko}, we have both
 the Hartree and the Fock diagrams in the two-loop order.
 This complicates the PMS analyses which will
 be reported elsewhere.  
\end{enumerate}

\subsection{Behavior of $m(T)$, $m_{0 \phi}(T)$ and $\xi(T)$}

 In Fig.\ref{vevm}(A) the tree-level masses in eq.(\ref{Rbmass})
 and  $m^2(T)$ are shown for $m_{\sigma}^{peak} (T=0)=550$ MeV. 
 $m_{0\phi}^2(T)$ is not tachyonic
 and approaches to $m^2(T)$ in the symmetric phase.
 This confirms that our resummation procedure cures the tachyon 
 problem in Sec.II.A.

 The solid line in Fig.\ref{vevm}(B) shows  the 
 chiral condensate $\xi(T)$ obtained by minimizing the
 free energy for the case $m_{\pi}(T=0) =140$ MeV, 
 with  $m_{\sigma}^{peak}(T=0) = 550$ MeV.
 $\xi(T)$ decreases uniformly as $T$ increases, which
 is a typical behavior of the chiral order parameter
 at finite $T$ away from the chiral limit.
 As we approach the chiral limit ($h \rightarrow 0$ or
 equivalently $m_{\pi} \rightarrow 0$), 
 $\xi(T)$ develops multiple solutions for given
 $T$, which could be an indication of the first
 order transition.  This will
 be discussed in more detail in the next subsection.
 The critical value of the quark mass $m_q^{\rm crit.}$
  below which the multiple solutions arise is
\beq
   m_q^{\rm  crit.} / m_q^{\rm phys.} = 
 (m_{\pi}^{\rm crit.}  / m_{\pi}^{\rm phys.})^2 = 0.08,
\eeq
where we have used Gell-Mann-Oakes-Renner relation \cite{GOR}
 to related the pion mass with the quark mass.
  $m_q^{\rm phys.}$ is the physical light-quark mass 
 corresponding to $m_{\pi}^{\rm phys.}=140$ MeV. 
 The critical temperature for 
   $ m_q^{\rm crit.} / m_q^{\rm phys.} = 0.08$ is
 $T_c \simeq 170$ MeV. 
 The behavior of $\xi(T)$ for $m_{\pi}(T=0)=30$ MeV (just below
 the critical value $m_{\pi}^{\rm crit.}$)
 is also shown by the
 dashed line in Fig.\ref{vevm}(B) for comparison.

 Fig.\ref{vevetc}(A)  shows $m_{0\phi}^2(T)$ for
 $m_{\sigma}^{peak} (T=0)=750,1000$ MeV. 
 The qualitative behaviors are similar to Fig. \ref{vevm}(A).
 The chiral condensate $\xi(T)$ for 
 $m_{\sigma} (T=0)=750,1000$ MeV is also shown in Fig.\ref{vevetc}(B).
 $\xi(T)$ is rather insensitive to the change of 
 $m_{\sigma}^{peak}$ as far as $m_{\pi}(T=0) =140$ MeV is imposed.

\subsection{Chiral limit ($h=0$)}

 In the chiral limit, the FAC condition and the resultant gap-equation
 are
 drastically simplified and some analytical study becomes
 possible.
 Let us carry out this analysis to reveal the nature of the
 chiral transition near the chiral limit.
 
  For $h=0$, the NG theorem is satisfied for
 given $m^2$ as  shown in Sec.II.E. Therefore, as far as
 $\xi \neq 0$ (the NG phase),
 the total self-energy of the pion must vanish
 at $(\omega, {\bf p}) = (0,{\bf 0})$:
\beq
\label{ng100}
m_{0\pi}^2 + \Sigma_{\pi}^R(0,{\bf 0};T) = 0 .
\eeq
A simultaneous solution of (\ref{ng100}) and
  the FAC condition (\ref{CHC}) is
 $m_{0\pi}^2 = \Sigma_{\pi}^R(0,{\bf 0};T) = 0$,
 which leads to
\beq
\label{sim2}
m^2 = - {\lambda \over 6} \xi^2 ,  \ \ 
m_{0 \pi}^2 = 0 , \ \ {\rm and} \ \ 
m_{0 \sigma}^2 = {\lambda \over 3} \xi^2 .
\eeq
The stationary condition (\ref{dvev}) has
 always a solution $\xi =0$ for $h = 0$ (the Wigner phase).
 The gap equation to determine
 the other solutions in the NG phase is obtained by
 substituting (\ref{sim2})
 into (\ref{dvev}):
\beq
\label{cgap}
0 & = &  \mu^2 + {\lambda \over 6} \xi^2  
  + {\lambda^2 \over 96 \pi^2} \xi^2
 \ln \left|{\lambda \xi^2 \over 3 \kappa^2 e} \right| \nonumber \\
 & &  + {\lambda \over 2} \int {d^3k \over (2\pi)^3}
       \left( {n_B(E_{\sigma}) \over E_{\sigma} } 
             + {n_B(E_{\pi}) \over E_{\pi} } \right) ,
\eeq
where $E_{\sigma}= \sqrt{ {\bf k}^2 + \lambda \xi^2 /3 }$ and
 $E_{\pi}= |{\bf k}|$. 

 The numerical solution of (\ref{cgap}) is given
 by the solid line
 in Fig.\ref{chiral}.  As can be seen from the figure,
 there are two non-vanishing solutions for $\xi$ in the range
 $T_{1}=126 \ {\rm MeV}  < T < T_{2}= 153 \ {\rm MeV} $, 
 which is a typical behavior of 
 the first order phase transition.
  For comparison,
  the case slightly away from the chiral limit is shown
 by the dashed lines in Fig.\ref{chiral}.
 
 $T_1$ and the behavior of $\xi(T)$  for $T \sim T_1$
 can be  solved analytically
 by expanding (\ref{cgap}) in terms of $\xi$
 near $\xi = 0$:
 Only the first term, $\mu^2$, and the last two terms
 proportional to the Bose-Einstein distribution
 are relevant, and one obtains
\beq
 T_1 = \sqrt{12 \over \lambda} | \mu | , \ \ 
\xi(T > T_1) \simeq {4 \pi \over \sqrt{3 \lambda}} (T-T_1) .
\eeq

 The existence of the multiple solutions of the gap equation
 for the
 $O(4)$ $\sigma$ model in the
 mean-field approach has been known for a long time
 \cite{first}.  Our analyses above confirm this feature
 within the framework of OPT. However,
 as is discussed in the second reference of \cite{first},
 this first order nature is likely to be
 an artifact of the mean-field approach, since
 the higher loops of massless $\pi$ and almost massless
 $\sigma$ are not negligible near $T_1$, and they 
 could easily change the order of the transition \cite{alsoo}.
 In fact, the renormalization group analyses as well as
 the direct numerical simulation on the lattice
 indicate that the $O(4)$ $\sigma$ model has a second order
 phase transition \cite{kanaya}.

 In the following, we will go back to the 
 the real world with $m_{\pi}(T=0)= 140 $MeV, where
 the gap-equation has only one solution for given $T$.

\subsection{Spectral function at $T \neq 0$}

 In Fig.\ref{spectT}(A),(B),  we show the spectral functions 
 $\rho_{\pi,\sigma}(\omega, {\bf 0};T)$
 for $T=50,120,145$ MeV with $m_{\sigma}^{peak} (T=0)=550$ MeV. 

 In  the $\pi$-channel,
 a continuum develops for  $ 0 < \omega < m_{0\sigma}-m_{0\pi}$.
 This originates from the induced ``decay'' by
 the scattering with thermal pions in the heat-bath;
 $\pi + \pi^{\rm thermal} \rightarrow \sigma $. Because of this 
 process, the pion acquires a width $ \sim 50 {\rm MeV}$
 at $T= 145$ MeV, while the peak position does not show 
 appreciable modification. They are
   in accordance with
 the Nambu-Goldstone nature of the pion, and are
 consistent with other calculations based on the 
 low $T$ expansion \cite{pwidth}. 

 In the $\sigma$-channel for $0 < T < 145 {\rm MeV}$, 
 there are two noticeable modification of the spectral function.
 One is the shift of the $\sigma$-peak toward the low mass region.
 The other is the sharpening of the
 spectral function just above the continuum threshold starting at
 $\omega = 2 m_{0\pi}(T)$.

 These features are simply controlled by  zeros
 or approximate zeros of 
\beq
\label{zeroo}
 {\rm Re} [D_{\sigma}^R(\omega,\vec{0})]^{-1}
 =  \omega^2 - m_{0\phi}^2 - {\rm Re} \Sigma^R_{\sigma}(\omega,\vec{0};T) ,
\eeq
 which appears in the denominator of the spectral function
 (\ref{spect2}). Note that the imaginary part
 of $[D_{\sigma}^R(\omega,\vec{0})]^{-1}$ is a smooth
 function of $\omega$ 
 and does not develop zeros above the threshold.
 Eq.(\ref{zeroo}) is plotted in Fig.\ref{spectT}(C). 
 For $T < 145 $ MeV, $ {\rm Re} [D_{\sigma}^R(\omega,\vec{0})]^{-1}$
 has only one zero for given $T$.  This zero corresponds to  an 
 ``effective'' mass of the $\sigma$-meson at finite $T$ and
 roughly corresponds to the position of the broad peak
  in Fig.\ref{spectT}(B). 
 
 On the other hand, as $T$ increases,  the cusp
  in the  low $\omega$ region starts to creat
 an  approximate zero of ${\rm Re} [D_{\sigma}^R(\omega,\vec{0})]^{-1}$.
 (At $T=145$ MeV, the cusp creates exact zero as shown
 in Fig.\ref{spectT}(C).)  
 This is why the peak just above the threshold 
 develops as $T$ increases as shown in Fig.\ref{spectT}(B). 
 The cusp originates from the
 $\sigma -\pi -\pi$ coupling (the fourth diagram for the
 sigma self-energy in  Fig.\ref{self}) and is related to the 
 the continuum threshold by analyticity.
 The position of the cusp is exactly the point
 where the continuum starts; $\omega = 2 m_{0\pi}(T)$.

 The approximate 
  shape of the spectral function for $\omega \simeq 2 m_{0\pi}(T)$ with
 $T \simeq 145$ MeV can be estimated as follows.
 The first term in the denominator of (\ref{spect2}) 
 approaches zero smoothly as
 $\omega \rightarrow 2 m_{0 \pi}$.
\beq
\label{1st}
 \left[ \omega^2 - m_{0\phi}^2 - 
 {\rm Re} \Sigma^R_{\sigma} (\omega,\vec{0})
\right]_{\omega \rightarrow 2 m_{0 \pi}}  \rightarrow 0 .
\eeq
 On the other hand, the imaginary part of the self-energy
 is a  
 phase space factor multiplied by  a smooth and non-zero 
 function $f$:
\beq
\label{2nd}
& &{\rm Im}\Sigma^R_{\sigma} (\omega,\vec{0};T) \nonumber \\
& & \ \ =  \theta(\omega - 2 m_{0\pi})
  \sqrt{1-{4m_{0\pi}^2 \over \omega^2} } f(\omega, T).
\eeq
 Substituting (\ref{1st}) and (\ref{2nd})
 into  eq.(\ref{spect2}), one finds
 \beq
& & \rho_{\sigma}(\omega \simeq 2 m_{0\pi},\vec{0};T) \nonumber \\ 
& & \ \ \ \ =  \theta( \omega - 2 m_{0\pi} ) {1 \over 
 \sqrt{ 1-  { 4m_{0\pi}^2 \over \omega^2} } f(2m_{0\pi}, T) }.
\eeq
 This explains the enhancement just above the threshold
 due to the phase space factor.

 There is another 
 explanation of the threshold enhancement.
 Let us start with  a sum rule for the spectral function which can be
 proved by the spectral decomposition of $\rho_{\phi}$:
\beq
\int_0^{\infty}  \rho_{\phi}(\omega,{\bf k};T)  d\omega^2 =
 (A+1)^{-1},
\eeq
 where $Z=A+1$ is the wave function renormalization constant
 and does not depend on $T$.
 Since $A$=0 in the one-loop order,
  the spectral integral  is unity for arbitrary $T$.
 This fact together with the positivity of
 $\rho_{\phi}$
 implies that there is a
  spectral concentration  near the threshold  as
 $m_{\sigma}^{peak}(T)$ decreases.  

 Beyond one-loop, $A$ is divergent in perturbation
 theory. However, one can always 
 define a  finite and $T$-independent 
 spectral integral as
\beq
 \int_0^{\infty} \left[ \rho_{\phi}(\omega,{\bf k};T)
  - \rho_{\phi}(\omega,{\bf k};T=0) \right] d\omega^2 =0.
\eeq
 Therefore, the same argument with the one-loop case holds
 and spectral concentration near threshold will
 occur even beyond one-loop.
 The threshold enhancement in Fig.\ref{spectT}(B), 
 although it occurs
 at relatively low $T$, is caused by a combined effect of the
 partial restoration of chiral symmetry (decreasing
 ``effective'' mass) and the strong $\sigma$-$2\pi$
 coupling. In the chiral limit, 
 the continuum threshold starts from $\omega=0$,
 and the enhancement occurs exactly at the critical temperature
 of  chiral transition. 

 Similar threshold enhancement in the $\pi$ channel
 becomes prominent  just below 
 $\omega = m_{0 \sigma}- m_{0 \pi}$ 
 for $T \simeq 165 $ MeV.  The basic mechanism
 of this enhancement is the same for the $\sigma$-case except that 
 ${\rm Re}[D^R_{\pi} (\omega,\vec{0})]^{-1}$ is a deceasing function
 of $T$.

 The spectral functions of $\pi$ ($\sigma$) at higher temperature
 exhibits the standard behavior  as expected from
 previous analyses \cite{HK85,huang}.
 Shown in Fig.\ref{psspectT} are simple
 $\sigma$ and $\pi$ poles and a continuum at $T=180$ MeV.
   As $T$ increases, these  poles 
 gradually merge into a degenerate (chiral symmetric) states.
 Due to this approximate degeneracy, 
 the normal decay of $\sigma$ 
 through $\sigma \rightarrow 2 \pi$ and the induced decay of $\pi$ 
 through $\pi + \pi \rightarrow \sigma$ are  kinematically forbidden
 at high $T$. This is why 
  the width of $\sigma$ and $\pi$ vanishes.

 For sufficiently high $T$,
 the system is supposed to be in the deconfined phase and the decay
 $(\sigma, \pi) \rightarrow q \bar{q}$ must occur. This 
 is not taken into account in the present linear $\sigma$ model.
 A calculation based on the Nambu-Jona-Lasinio model shows, however, that
 there is still a chance for collective modes to survive
 as far as  $T/T_c$ is not so far from unity \cite{HK85}.

\subsection{Diphoton emission rate through $\sigma \rightarrow 2 \gamma$}

 As one of the experimental candidates to see the 
 threshold enhancement in the $\sigma$ channel,
 we evaluated the diphoton emission rate from the decay
 $\sigma \rightarrow 2 \gamma$ in hot hadronic matter  \cite{chiku97}.
 The diphoton yield (with back to back kinematics) per unit space-time 
 volume of a hot hadronic plasma  can be 
 written as \cite{kapusta}
\beq
\label{rate}
  {dR_{\sigma} \over d^4 x d^4 k} = {1 \over (2 \pi)^4} 
  g_{\sigma}^2 |F_{\sigma \gamma \gamma}(k^2)|^2 \omega ^4
  { \rho_{\sigma}(\omega,{\bf k} =0;T) \over e^{ \omega /T}-1},
\eeq
 where $k^{\mu}=(\omega , {\bf k})$ denotes the total four-momentum of 
 the diphoton.
 $g_{\sigma}$ is the
 $\sigma \gamma \gamma$ vertex at zero momentum
 and  $F_{\sigma \gamma \gamma}(k^2)$ is a corresponding form
 factor.
 $ g_{\sigma} F_{\sigma \gamma \gamma}$
 has a short  distant contribution from the constituent-quark loop 
 and a long distant contribution
 from the pion-loop. We took the estimates 
 given in \cite{gsgg} for these contributions.
 Formula for $\pi^0 \rightarrow 2 \gamma$ is obtained by
 a replacement   $\sigma \rightarrow 2 \gamma$ in (\ref{rate}).
 In this case, $g_{\pi}$ is fixed by the axial anomaly, and 
 $ F_{\pi \gamma \gamma}$ is taken from an estimate
 using the  chiral quark model \cite{gmgg}.

 The main background for the above processes is
 the thermal annihilation of pions; $\pi^+ \pi^- \rightarrow 2 \gamma$.
 Diphoton yield from $\sigma \rightarrow 2 \gamma$ 
 and $\pi^0 \rightarrow 2 \gamma$ 
 together with this background at T=145 MeV are shown 
 in Fig.\ref{2gamma}.
 Threshold enhancement in $\sigma \rightarrow 2 \gamma$
 process is significant only 
 in a narrow region of 
 the diphoton invariant mass
 and in a narrow region of $T$.
 Similar conclusion is drawn in 
 other analysis \cite{reh}.

\section{Summary}

 In this paper, we have examined the
 optimized perturbation theory (OPT) in detail  at finite $T$.
 For theories with spontaneous symmetry breaking,
 the loop-wise expansion in 
 OPT is shown to be a suitable scheme to resum higher order terms.

 We have shown that OPT naturally cure the two major problems of the
  naive loop-expansion, namely 
 breakdown of perturbation series at high $T > T_c$ and 
 the existence of tachyon poles for  $T < T_c$.

 We have also shown that 
 OPT has several advantages over  
 other resummation methods proposed so far.
 First of all, the renormalization of the UV
 divergences, which is not a trivial issue in other 
 methods, can be carried out systematically 
 in  the loop-expansion in OPT. 
 This is because one can separate the 
 the self-consistent procedure (Step 3 in Sec.II.B)
 from the renormalization procedure (Step 2 in Sec.II.B) in OPT.

 Whether the Nambu-Goldstone (NG) theorem
 is fulfilled in  resummation methods has been
 discussed in the literatures.
 We found  that the loop-expansion in OPT can give a clear
 view on this problem.
 The NG theorem is a direct consequence of the
 invariance of the effective action.  Since OPT presented
 in this paper does not break the global symmetry
 of the effective potential 
  in each order of the perturbation,
  one can  prove, without much difficulty,
 that the NG theorem is fulfilled in any give order
 of the loop-expansion in OPT.

 In the latter part of this paper, we 
 have applied OPT to the $O(4)$ $\sigma$ model to study
 the spectral functions 
 at finite $T$.  The OPT in one-loop order
 together with a FAC condition for the pion 
 self-energy, we have successfully summed not only
 the cactus diagrams but also  other loop diagrams.
 We have demonstrated  that the spectral function of $\sigma$, 
 which does not show a clear resonance at $T=0$, develops
 a sharp enhancement near 2$\pi$ threshold as $T$ approaches
 $T_c$. This is due to a combined effect
 of the partial restoration of chiral symmetry and 
 the strong $\sigma - 2 \pi$ coupling. Although it is rather
 difficult to observe this enhancement in the 
 diphoton spectrum,
 further studies will be necessary to reveal the phenomenological
 implications of this phenomena.

 The basic idea of OPT examined in detail in this paper
 will also have relevance to develop an improved perturbation
 theory for gauge theories in which the 
 weak coupling expansion is known to break down in high orders \cite{linde80}.
 A generalization of OPT, such as that
 discussed at Step 2 in  Sec. II.B,  will be necessary
 for this purpose.

\section*{Acknowledgments}

The authors would like to thank M. Asakawa, 
K. Kanaya, T. Kunihiro, T. Matsui and J. Randrup
  for useful discussions.
This work was partially supported by
the Grants-in-Aid of the Japanese Ministry of 
Education, Science and Culture (No. 06102004).
S. C. would like to thank the Japan Society of Promotion of Science (JSPS)
for financial support.

\onecolumn

\appendix

\section{Counter terms in OPT}

 Consider the Lagrangian ${\cal L}(\phi;m^2)$ and 
 define $\Gamma^{(n,j)}_R (\lambda, m^2)$ as the 
 the renormalized $n$-point proper vertex with insertion
 of the composite operator $\phi^2(x)/2$ by $j$-times.
 (We do not write the
 external momentums explicitly.)
 Applying the counter terms in (\ref{count2}), 
 $\Gamma^{(n,j)}_R$
 is written by the unrenormalized
 proper vertex $\Gamma^{(n,j)}_0$ and the renormalization
 constants ($Z$, 
 $Z_{\phi^2}$, $\Delta_1$ and $\Delta_2$) as \cite{IZ}
\beq
\label{a1}
 \Gamma^{(n,j)}_R (\lambda, m^2) =
 Z^{n/2} Z_{\phi^2}^{-j} 
 \Gamma^{(n,j)}_0(\lambda_0,m^2_0)
 +  (2 \Delta_2 \delta_{j2}  + \Delta_1 \delta_{j1} ) \delta_{n0} ,
\eeq
where the bare quantities with a  suffix ``0''
 are related to the renormalized quantities as
 $\lambda_0 = Z_{\lambda} \lambda$,
  $m_0^2 = Z_{\mu} m^2$ and $\phi_0 = \sqrt{Z} \phi$.

 To show (\ref{zet1}), 
 consider the proper self-energy  $\Gamma^{(2,0)}_R(\lambda,m^2)$
 and its derivative with respect to $m^2$. After 
 a straightforward algebra using (\ref{a1}), one finds
\beq
\label{zet11}
 {\d  \over \d m^2 } \Gamma^{(2,0)}_R(\lambda,m^2)
 = - Z Z_{\mu} \Gamma^{(2,1)}_0(\lambda_0,m^2_0) 
 = - Z_{\mu} Z_{\phi^2} \Gamma^{(2,1)}_R(\lambda,m^2).
\eeq
 Since $\Gamma^{(n,j)}_R$ is finite,
 eq.(\ref{zet11}) implies 
  $Z_{\mu} Z_{\phi^2}$ must be  finite
 in four dimensions. 
 Now, in the loop-expansion with the 
 $\overline{MS}$ scheme, $Z_{\mu}$ and $Z_{\phi^2}$ 
 have expansions of a form
   $1 + \sum_{l=1}^{\infty} a_l \delta^l$
 with $a_l$ containing only the powers of $1/\bar{\vep}$.
 This fact together with the  finiteness of  $Z_{\mu} Z_{\phi^2}$ 
 for $\bar{\vep} \rightarrow 0$
 immediately leads to eq.(\ref{zet1}),   $Z_{\mu} Z_{\phi^2}=1$.

 To show (\ref{zet2}), 
 consider the proper vertex without 
 external legs:  $\Gamma^{(0,0)}_R(\lambda,m^2)$.
 This quantity is a sum of all the one-particle irreducible
 diagrams +  the $c$-number counter term $Dm^4$.  
 Therefore, taking a derivative with respect to $m^2$ and
 using (\ref{a1}), one finds
\beq
\label{zet22}
 {\d  \over \d m^2 } \Gamma^{(0,0)}_R(\lambda,m^2)
  =  -  Z Z_{\mu} {1 \over V_4} \int_{V_4} d^4x \ 
 \langle {\phi^2(x) \over 2} \rangle_{1PI}
 + 2 Dm^2 
  =  -[ \Gamma^{(0,1)}_R(\lambda,m^2) - \Delta_1 ] + 2Dm^2.
\eeq
 Here $\langle \cdot \rangle_{1PI}$ denotes the
 one-particle irreducible contribution. $V_4$ denotes the
 Euclidean four-volume: At $T \neq 0$, $V_4 = V_3/T$
 with $V_3$ being the three-volume.
 Since $\Gamma^{(n,j)}_R$ is finite, eq.(\ref{zet22}) implies that
 $2Dm^2 + \Delta_1$ must be finite in
 four dimensions. $D$ and $\Delta_1$ have expansions
 of the form, $\sum_{l=1}^{\infty} a_l \delta^l$,
 with $a_l$ containing only the powers of $1/\bar{\vep}$.
 Thus one arrives at eq.(\ref{zet2}), $2Dm^2 + \Delta_1 =0$.

 To show (\ref{zet3}), 
 one needs to take derivative with respect to $m^2$ one more time:
\beq
\label{zet33}
\left( {\d  \over \d m^2 } \right) ^2 \Gamma^{(0,0)}_R(\lambda,m^2)
 =  Z^2 Z_{\mu}^2 {1 \over V_4} \int_{V_4} d^4x \int_{V_4} d^4y
  \ \langle {\phi^2(x) \over 2}
  {\phi^2(y) \over 2} \rangle_{1PI}
 + 2 D 
 = [ \Gamma^{(0,2)}_R(\lambda,m^2) - 2 \Delta_2 ] + 2D.
\eeq
 By the similar argument as above, 
 $D - \Delta_2 =0 $ follows from eq.(\ref{zet33}).

\section{One-loop formula for self-energy at $T\neq 0$}
\begin{figure}[h]
%  \begin{center}
\centerline{
    \epsfxsize=18.0cm
     \epsfbox{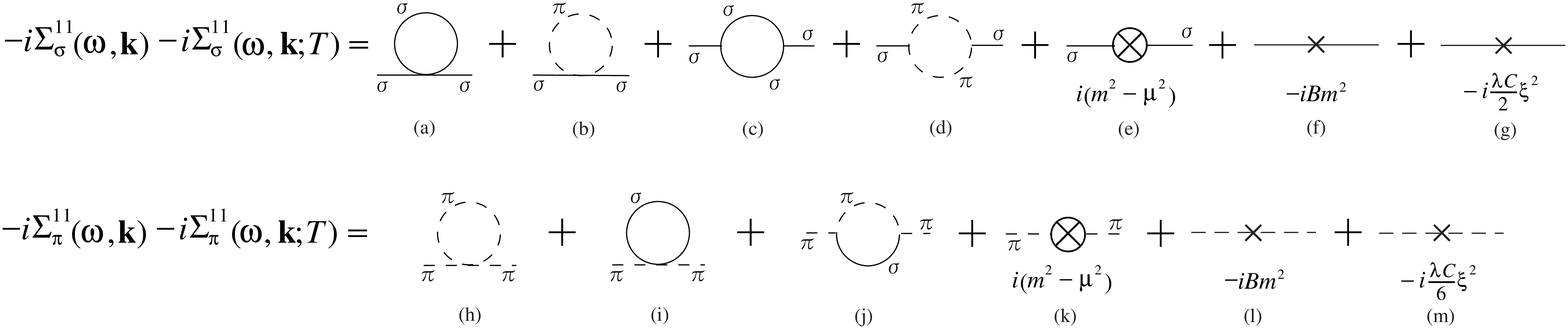}
}
%  \end{center}
     \caption{One-loop self-energy  $\Sigma^{11}$ for $\sigma$ and 
     $\pi$ in the  modified loop-expansion at finite $T$.}
 \label{self}
\end{figure}

\begin{figure}[h]
%  \begin{center}
\centerline{
    \epsfysize=6.3cm
    \epsfbox{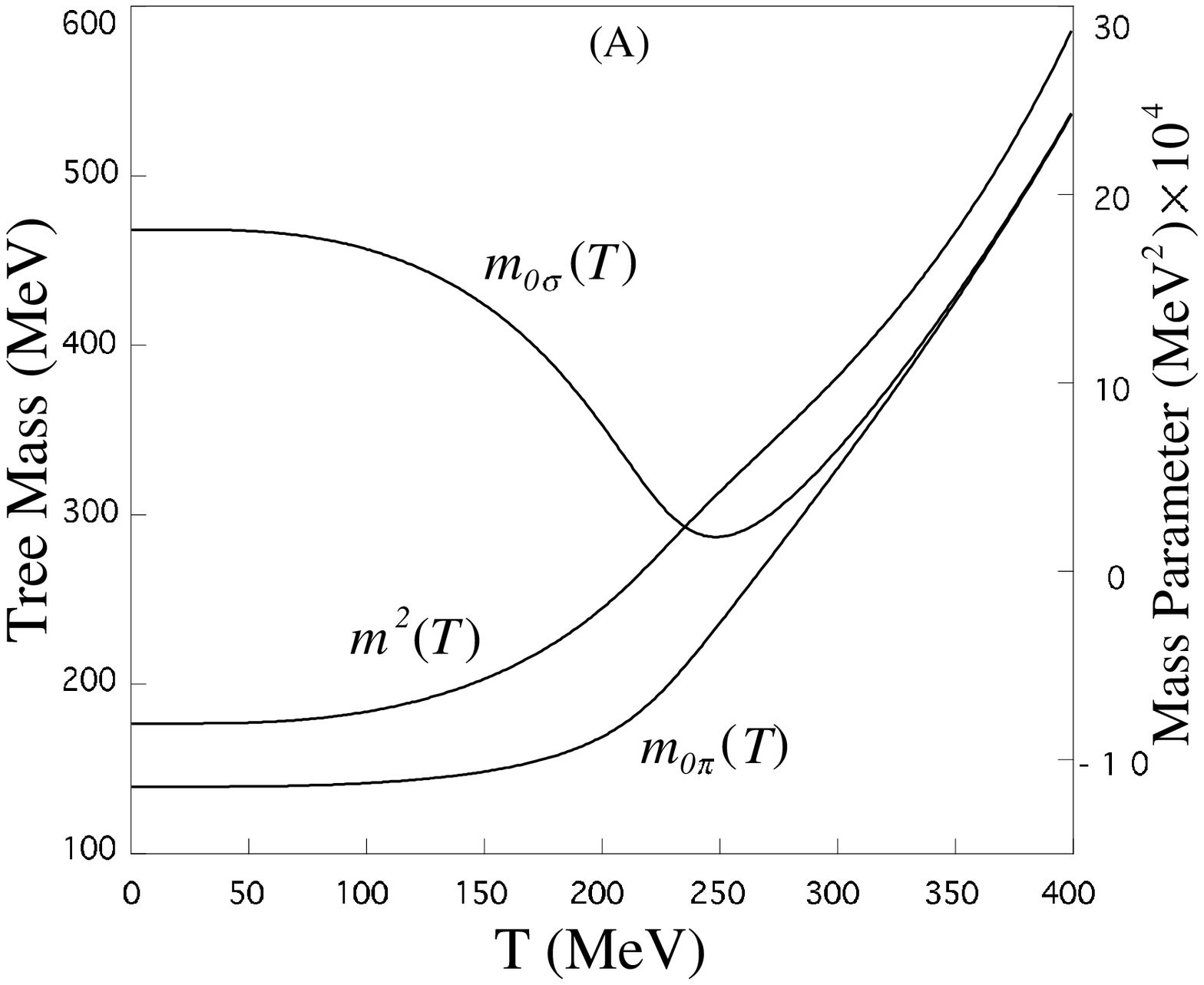}\hspace{0.2cm}
    \epsfysize=6.3cm
    \epsfbox{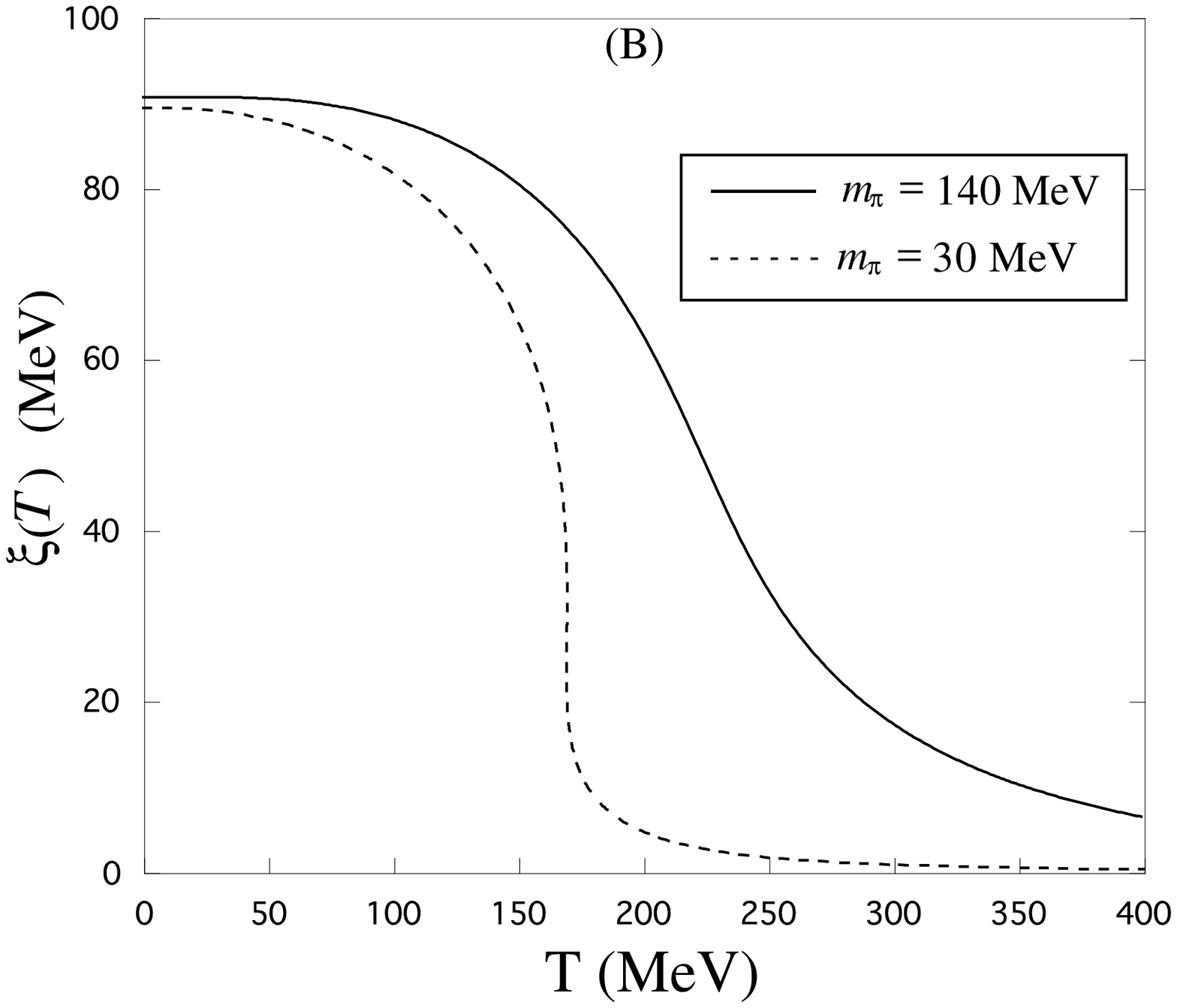}
}
%  \end{center}
     \caption{
     (A) Masses in the tree-level $\mp(T)$ and 
     $\ms(T)$ shown with left vertical scale, and 
     the mass parameter $m^2(T)$ with the right vertical scale.
     (B) $\xi(T)$ for $m_{\pi}(T=0)=140 $MeV and 30 MeV with 
     $m^{peak}_{\sigma}(T=0)=550$ MeV.}
 \label{vevm}
\end{figure}

\begin{figure}[h]
%  \begin{center}
\centerline{
    \epsfysize=6.3cm
    \epsfbox{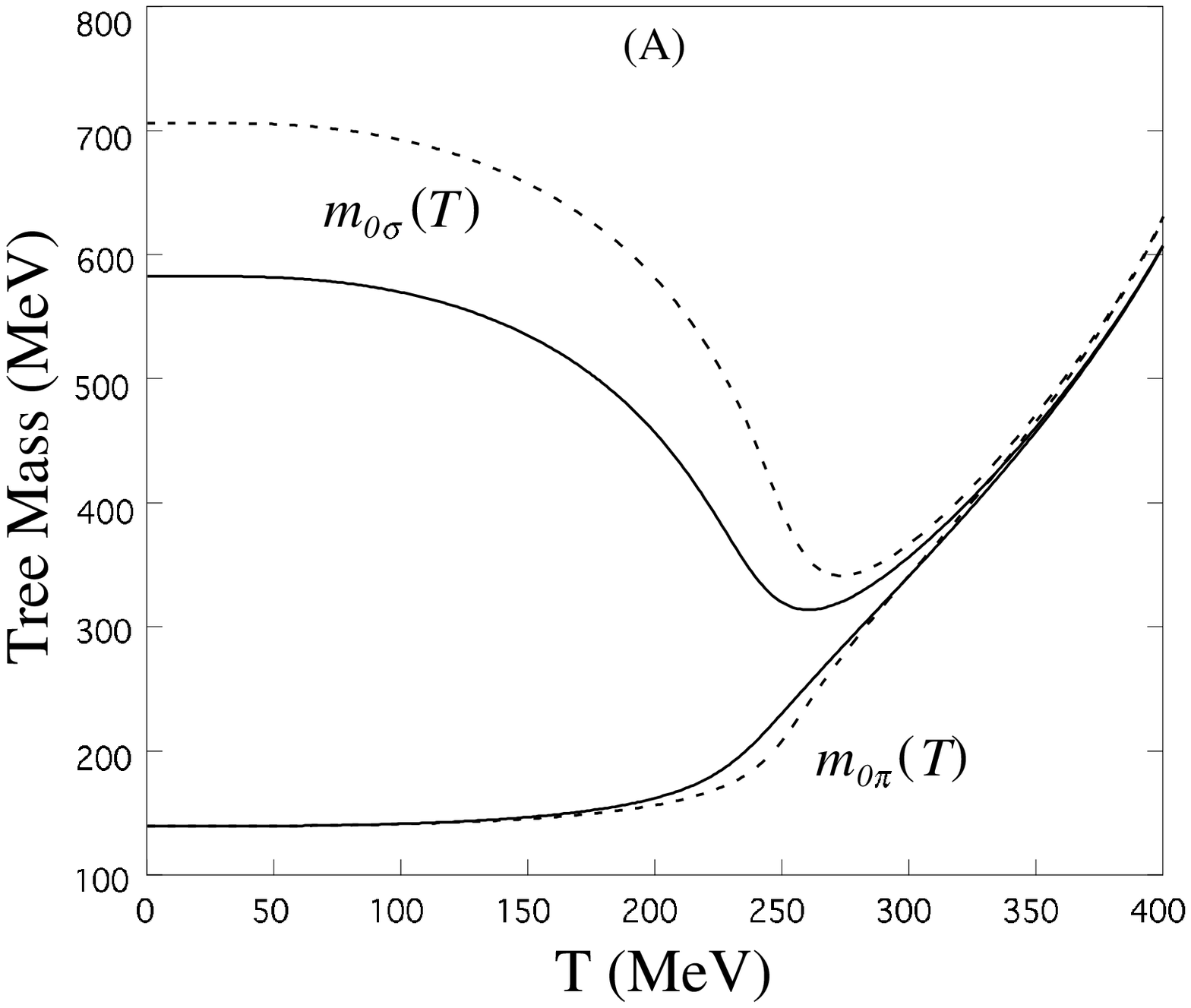}\hspace{0.2cm}
    \epsfysize=6.3cm
    \epsfbox{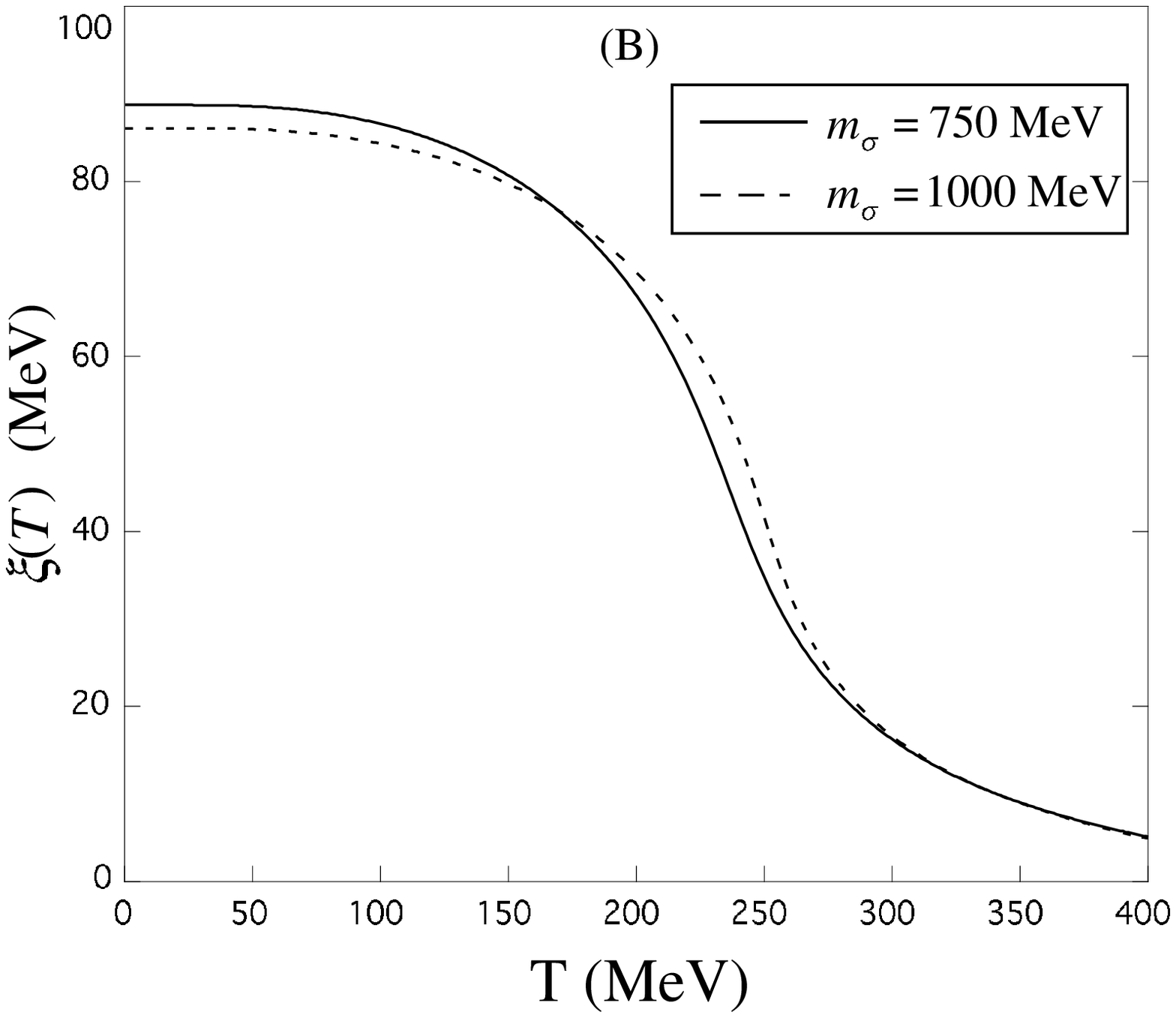}
}
%  \end{center}
     \caption{(A) Masses in the tree-level $\mp(T)$ and 
     $\ms(T)$. (B) $\xi(T)$ for $m^{peak}_{\sigma}(T=0)
    =750 $MeV and 1000 MeV
     with $m_{\pi} =140$ MeV.}
 \label{vevetc}
\end{figure}

\begin{figure}[h]
%  \begin{center}
\centerline{
    \epsfysize=6.3cm
    \epsfbox{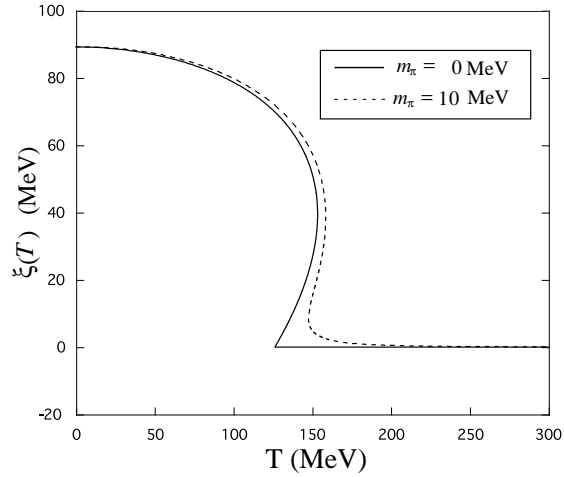}\hspace{0.2cm}
}
%  \end{center}
     \caption{
      $\xi(T)$ for $m^{peak}_{\sigma}(T=0)
    = 550 $MeV with $m_{\pi} =0$ MeV and $m_{\pi}=10$ MeV.}
 \label{chiral}
\end{figure}

\begin{figure}[h]
%  \begin{center}
\centerline{
    \epsfysize=6.3cm
    \epsfbox{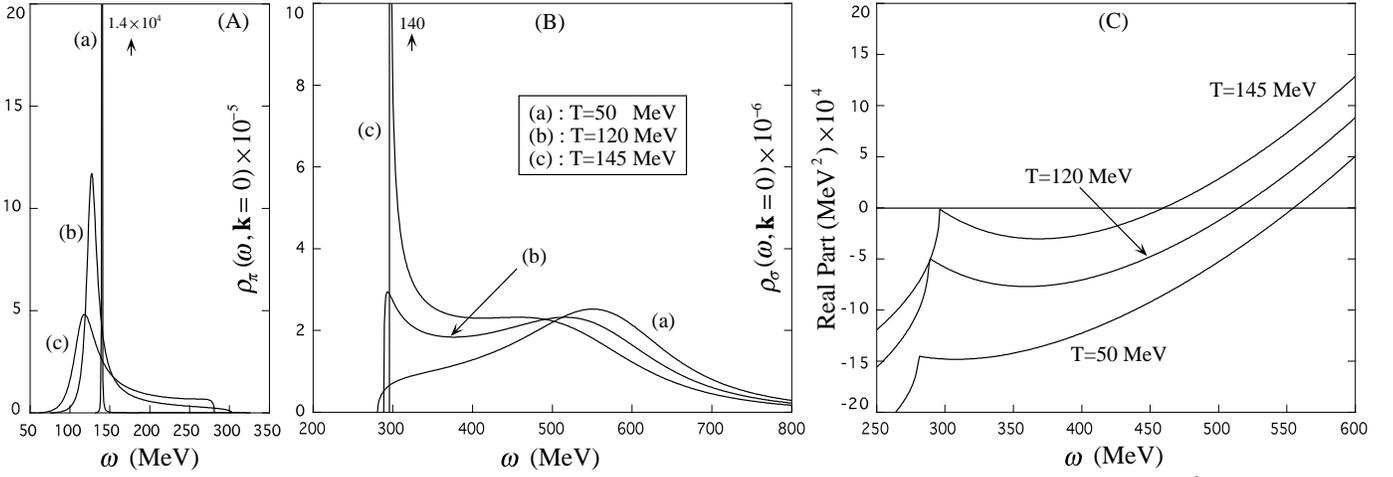}
}
%  \end{center}
     \caption{Spectral function in the 
     $\pi$ channel (A) and in the $\sigma$ channel
     (B) for $T=50, 120, 145 $ MeV with $m^{peak}_{\sigma} (T=0)=550$ MeV.
     The real part of $(D_{\sigma}^R(\omega, 0;T))^{-1}$
     as a function of $\omega$ is shown in (C).}
 \label{spectT}
\end{figure}

\begin{figure}[h]
%  \begin{center}
\centerline{
    \epsfysize=6.3cm
    \epsfbox{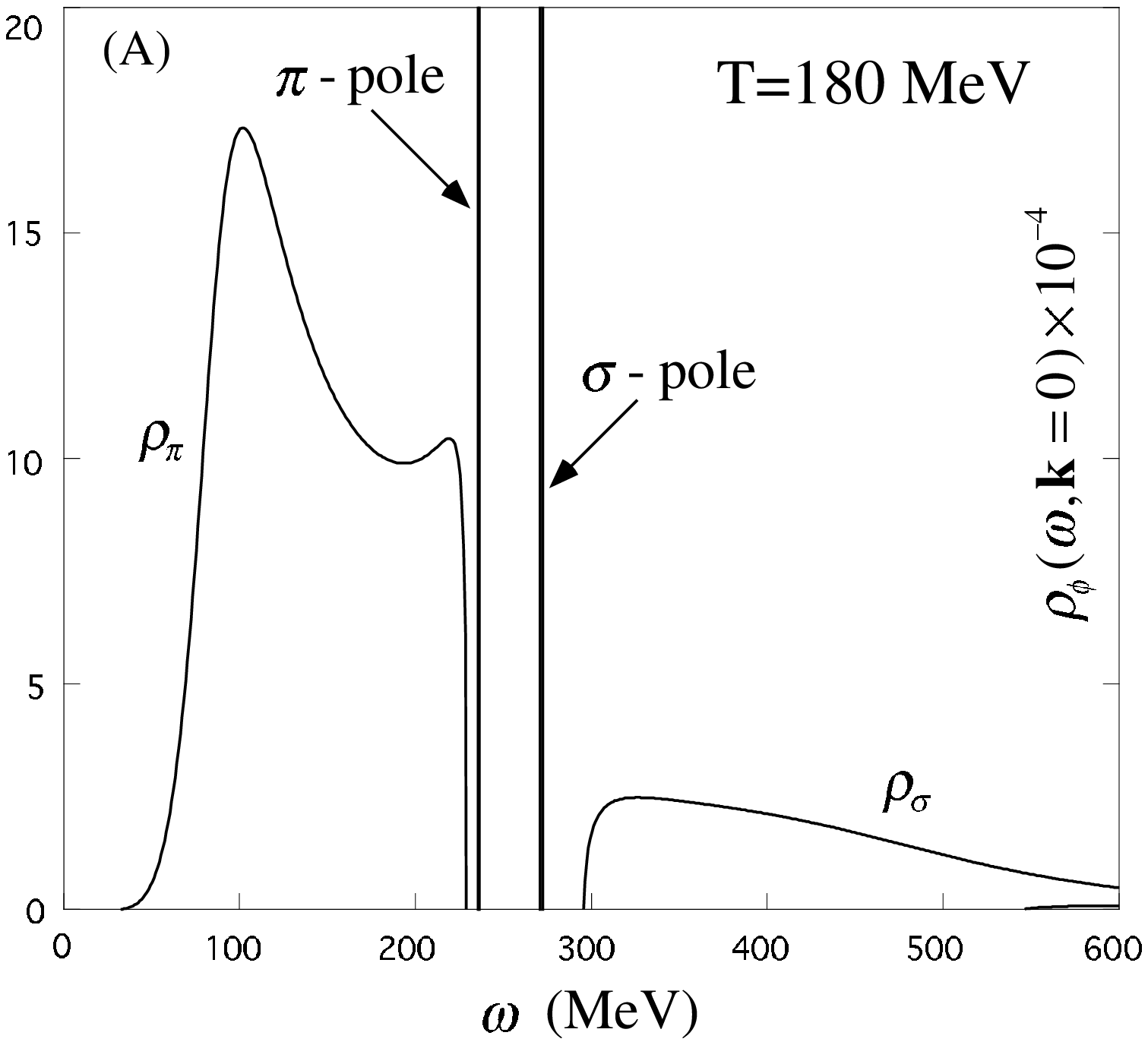} \hspace{0.2cm}
    \epsfysize=6.3cm
    \epsfbox{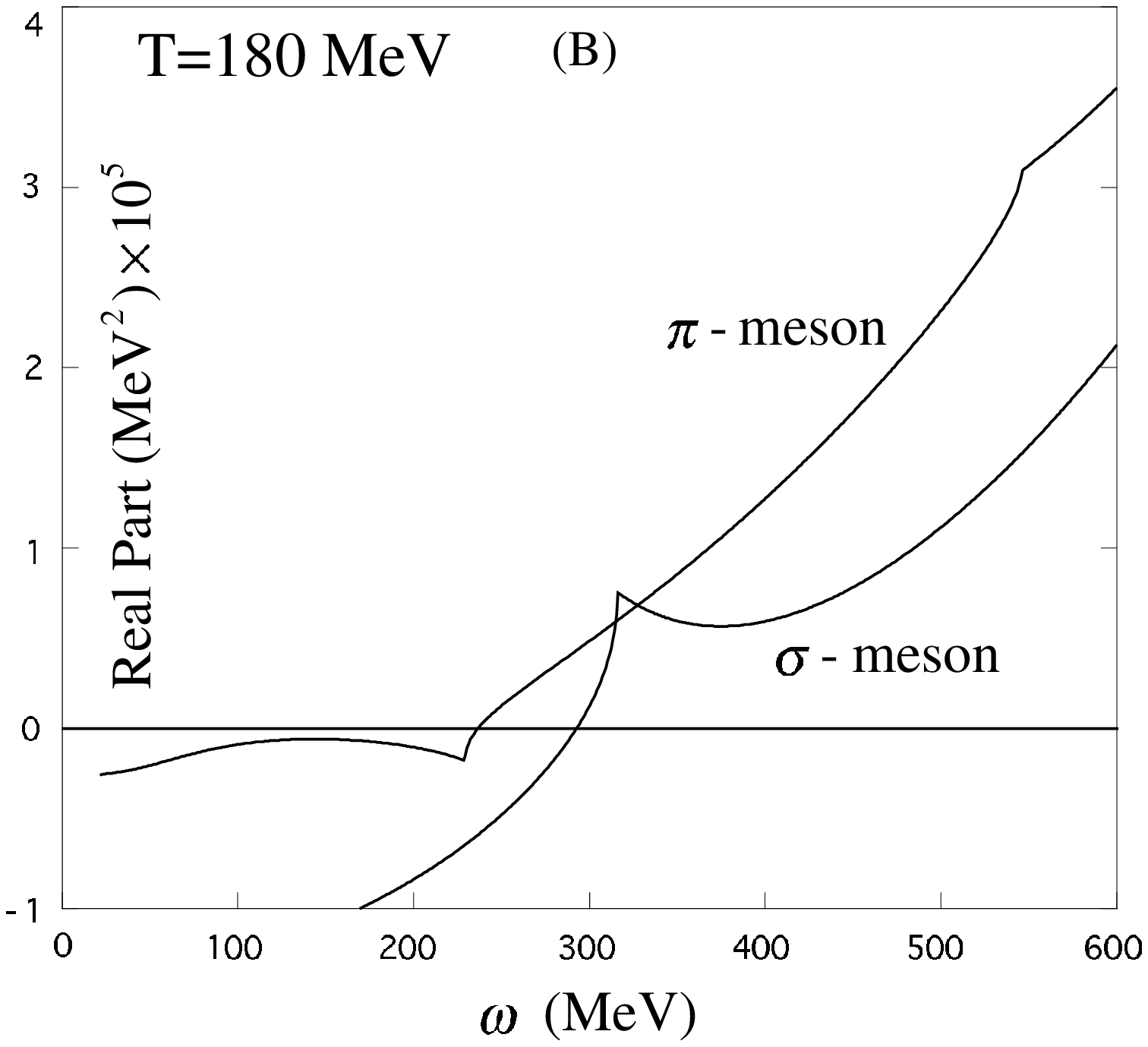}
}
%  \end{center}
     \caption{(A) 
      Spectral functions in the $\pi$ and 
      $\sigma$ channels at $T=180$ MeV.
      (B) The real part of $(D_{\pi}^R(\omega, 0;T))^{-1}$
      and $(D_{\sigma}^R(\omega, 0;T))^{-1}$ as a function 
      of $\omega$ at $T=180$ MeV
      with $m^{peak}_{\sigma} (T=0)=550$ MeV.}
 \label{psspectT}
\end{figure}

\begin{figure}[h]
%  \begin{center}
\centerline{
    \epsfysize=6.3cm
    \epsfbox{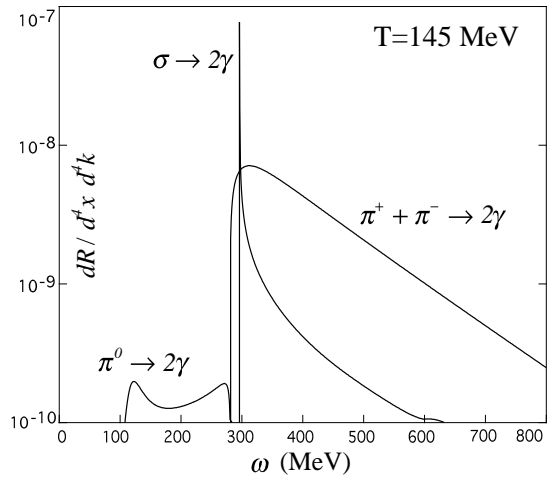}
}
%  \end{center}
     \caption{Diphoton yield per unit space-time volume in the 
     back to back
     kinematics at T=145 MeV for $m^{peak}_{\sigma} (T=0)=550$ MeV.}
\label{2gamma}
\end{figure}

Formulas corresponding to Fig.\ref{self} read
\beq
  -i \Sigma^{11}_{\sigma}( \omega,{\bf k})
  -i \Sigma^{11}_{\sigma}( \omega,{\bf k};T) & = &
  -i {\lambda  \over 2} [ I_{\pi}^{(1)}+F_{\pi}^{(1)}+ 
  I_{\sigma}^{(1)}+F_{\sigma}^{(1)}]
  +(-i {\lambda  \xi \over 3})^2 
  {3 \over 2}[I_{\pi}^{(2)}+2F_{\pi}^{(2)}+F_{\pi}^{(3)}] \\
  && \mbox{}+(-i \lambda  \xi)^2 {1 \over 2} 
  [I_{\sigma}^{(2)}+2F_{\sigma}^{(2)}
  +F_{\sigma}^{(3)}] + i(m^2- \mu^2) + {\rm counter \ terms} \nonumber \\
  -i \Sigma^{11}_{ \pi}( \omega ,{\bf k})
  -i \Sigma^{11}_{\pi}( \omega,{\bf k};T) & = &
  -i {5 \lambda  \over 6} [ I_{\pi}^{(1)}+F_{\pi}^{(1)}]
  -i {\lambda  \over 6} [ I_{\sigma}^{(1)}+F_{\sigma}^{(1)}] \\
  && \mbox{}+(-i {\lambda  \xi \over 3})^2 
  [I^{(3)}+F^{(4)}+F^{(5)}]
  + i(m^2- \mu^2) + {\rm counter \ terms}, \nonumber
\eeq
with
\beq
\label{I1}
  I_{ \phi}^{(1)} & = & \kappa^{2 \vep} \int {d^4 p \over (2 \pi)^4} 
  {i \over p^2-m_{0 \phi}^2+i\epsilon}, \\
\label{I2}
  I_{ \phi}^{(2)} & = & \kappa^{2 \vep} \int {d^4 p \over (2 \pi)^4} 
  {i \over p^2-m_{0 \phi}^2+i\vep}{i \over (p+k)^2-m_{0 \phi}^2+i\epsilon}, \\
\label{I3}
  I^{(3)} & = & \kappa^{2 \vep} \int {d^4 p \over (2 \pi)^4}
  {i \over p^2-m_{0 \sigma}^2+i\vep}{i \over (p+k)^2-m_{0 \pi}^2+i\epsilon}, \\
\label{F1}
  F_{\phi}^{(1)} & = &\int \frac{d^4p}{(2\pi)^4}2\pi n_{B}(|p_0|)
  \delta(p^2-\mh^2),\\
\label{F2}
  F_{\phi}^{(2)}& = &i \int \frac{d^4p}{(2\pi)^4}\frac{2\pi n_{B}(|p_0|)
  \delta(p^2-\mh^2)}{(p+k)^2-\mh^2+i\epsilon},\\
\label{F3}
  F_{\phi}^{(3)}& = &\int \frac{d^4p}{(2\pi)^4} (2\pi)^2n_{B}(|p_0|)n_{B}
  (|p_0+\omega |)\delta(p^2-\mh^2)\delta((p+k)^2-\mh^2),\\
\label{F4}
  F^{(4)} & = & i \int \frac{d^4p}{(2\pi)^4} \frac{2\pi n_{B}(|p_0|)
  \delta(p^2-\ms^2)}{(p+k)^2-\mp^2+i\epsilon} 
  + ( \ms \leftrightarrow \mp),\\
\label{F5}
  F^{(5)} & = &\int \frac{d^4p}{(2\pi)^4} (2\pi)^2n_{B}(|p_0|)n_{B}
  (|p_0+\omega |)\delta(p^2-\ms^2)\delta((p+k)^2-\mp^2).
\eeq
Here   $\vep = (4-n)/2$,
 $p^2 = p_0^2 - {\bf p}^2$, $k^2 = \omega^2 - {\bf k}^2$
 and  $n_{B}( \omega)=[e^{\omega /T}-1]^{-1}$.

The explicit forms of eq.(\ref{I1}), (\ref{I2}), (\ref{I3}) 
 for $k^2 > 0$ are
\beq
  I_{ \phi}^{(1)} & = & -{ \mh^2 \over 16 \pi^2}
  (\frac{1}{\bar{\vep}}+1-\log {\mh^2 \over \kappa^2}),  \\
  I_{ \phi}^{(2)} & = & -i{1 \over 16\pi^2} 
  [ \frac{1}{\bar{\vep}}-\log{\mh^2 \over \kappa^2}+2
   + \left\{ \begin{array}{lcl}
  q_2 (\log{1-q_2 \over 1+q_2 }+i\pi) ] & \mbox{for} & k^2 > 4\mh^2 \\[0.1cm]
  -2q_2 \arctan{1 \over q_2 } ] & \mbox{for} & 0 < k^2 < 4\mh^2, \end{array}
  \right. \\
  I^{(3)} & = & -i{1 \over 16 \pi^2}
  [ \frac{1}{\bar{\vep}}- \log{\mp^2 \over \kappa^2} +2
  +\frac{k^2+\ms^2-\mp^2}{2k^2} \log \frac{\mp^2}{\ms^2} \\
& & \mbox{} - \left\{ \begin{array}{lcl}
  q_3 (\log \frac{(2k^2 q_3 +k^2-\ms^2+\mp^2)(2k^2 q_3 +k^2+\ms^2-\mp^2)}
  {(2k^2 q_3 -k^2+\ms^2-\mp^2)(2k^2 q_3 -k^2-\ms^2+\mp^2)}-2i\pi)]
  & \mbox{for} & (\ms+\mp)^2<k^2 \\[0.2cm]
  2q_3 (\arctan \frac{k^2-\ms^2+\mp^2}{2k^2q_3}+
  \arctan \frac{k^2+\ms^2-\mp^2}{2k^2q_3})]
  & \mbox{for} & (\ms-\mp)^2<k^2<(\ms+\mp)^2\\[0.2cm]
  q_3 \log \frac{(2k^2 q_3 +k^2-\ms^2+\mp^2)(2k^2 q_3 +k^2+\ms^2-\mp^2)}
  {(2k^2 q_3 -k^2+\ms^2-\mp^2)(2k^2 q_3 -k^2-\ms^2+\mp^2)} ]
  & \mbox{for} & 0<k^2<(\ms-\mp)^2, \end{array} \right.  \nonumber
\eeq
with
\beq
  q_2  =  \sqrt{\left| \frac{4\mh^2}{k^2}-1 \right|}, \ \ \ \ 
  q_3  =  \frac{\sqrt{|(k^2+\ms^2-\mp^2)^2-4k^2\ms^2|}}{2k^2}. \nonumber
\eeq

 Eq.(\ref{F1}), (\ref{F2}), (\ref{F3}), (\ref{F4}) and (\ref{F5}) 
 for ${\bf k}=0$ read
\beq
  F_{\phi}^{(1)} & = &\int_0^{\infty} \frac{dp}{2\pi^2} \frac{p^2n_{B}(E(\mh))}
  {E(\mh)},\\
  F_{\phi}^{(2)} 
& = & i \int_0^{\infty} \frac{dp}{2\pi^2} \frac{p^2n_{B}(E(\mh))}
  {E(\mh)}\frac{1}{\omega^2-4E^2(\mh)} +
     \theta(\omega^2-4 \mh^2)
     \frac{\sqrt{\omega^2-4 \mh^2}}{16\pi \omega}n_{B}({\omega \over 2}), \\
  F_{\phi}^{(3)} & = &
     \theta(\omega^2-4 \mh^2)
  \frac{\sqrt{\omega^2-4 \mh^2}}{8\pi \omega}n_{B}^2({\omega \over 2}), \\
  F^{(4)} & = & i \int_0^{\infty} \frac{dp}{(2\pi)^2} \frac{p^2n_{B}(E(\ms))}
  {E(\ms)}\{\frac{1}{(\omega+E(\ms))^2-E(\mp)^2}+\frac{1}
  {(\omega-E(\ms))^2-E(\mp)^2}\} \nonumber\\
      & &\hspace{1cm}+\left\{\begin{array}{l}
      \frac{1}{16 \pi \omega^2}\sqrt{( \omega^2+\ms^2-\mp^2)^2-4 \ms^2 
      \omega^2}\ n_{B}(\frac{| \omega^2+\ms^2-\mp^2|}{2 \omega})\\[0.1cm]
      \hspace{3.2cm}\mbox{for}\ \ \ \ 0< \omega^2<(\ms-\mp)^2\ ,
      \ (\ms+\mp)^2< \omega^2\\[0.2cm]
      0\hspace{3cm}  \mbox{for}\ \ \ \  (\ms-\mp)^2 < \omega^2<(\ms+\mp)^2,
  \end{array}\right.\\[0.2cm]
  & &  + ( \ms \leftrightarrow \mp) \nonumber\\[0.1cm]
  F^{(5)} & = &\left\{\begin{array}{l}
  \frac{1}{8\pi\omega^2}\sqrt{(\omega^2+\ms^2-\mp^2)^2-4 \ms^2 \omega^2}\ n_{B}
  (\frac{| \omega^2+\ms^2-\mp^2|}{2 \omega})\ n_{B}
  (\frac{| \omega^2-\ms^2+\mp^2|}{2 \omega})\\[0.1cm]
  \hspace{4.6cm} \mbox{for}\ \ \ \ 0< \omega^2<(\ms-\mp)^2\ ,
  \ (\ms+\mp)^2< \omega^2\\[0.2cm]
  0\hspace{4.4cm}  \mbox{for}\ \ \ \  (\ms-\mp)^2 < \omega^2< (\ms+\mp)^2,
  \end{array}\right.
\eeq
 where $E(m)=\sqrt{{\bf p}^2+m^2}$.

\twocolumn

\onecolumn
%
%%%%%%%%%%%%%%%
%  Table
%%%%%%%%%%%%%%%
%
\begin{table}
\[ 
\begin{array}{|c||c|c|c|c|c|} \hline 
  \ m_{\sigma}^{peak} \ (\mev) 
& \mu^2 \ (\mev^2) \ & \  \lambda  & h \ (\mev^3) \ &
  \  \kappa \ (\mev) \ & \  \Gamma \ (\mev) \ \\ \hline
  550 & -284^2 & 73.0 & 123^3 & 255 & 260 \\
  750 & -375^2 & 122  & 124^3 & 325 & 657\\
 1000 & -469^2 & 194  & 125^3 & 401 & 995 \\ \hline
\end{array} 
\]
\caption{Vacuum parameters corresponding to $m_{ \sigma}^{peak}
 =$ 550, 750, 1000 MeV} 
\label{tab1}
\end{table}
%%%%%%%%%%%%%%%%%
%   eps files 
%%%%%%%%%%%%%%%%%
%
\end{document}